\documentclass[12pt]{article}
\usepackage[letterpaper,margin=1in]{geometry}
\usepackage{setspace}
\usepackage{amsmath,amssymb,amsthm,bm,bbm}
\usepackage{graphicx,booktabs,longtable}
\usepackage[ruled,lined]{algorithm2e}
\usepackage{placeins}
\usepackage{natbib}
\usepackage[hypertexnames=false]{hyperref}
\hypersetup{colorlinks=true,linkcolor=blue,citecolor=blue,urlcolor=blue}
\newtheorem{prop}{Proposition}
\newtheorem{lemma}{Lemma}
\theoremstyle{definition}
\newtheorem{ass}{Assumption}
\newtheorem{remark}{Remark}

\newcommand{\given}{\,|\,}
\newcommand{\I}{\mathbbm{1}}
\newcommand{\E}{\mathbb{E}}
\newcommand{\prob}{\mathbb{P}}
\newcommand{\V}{\operatorname{\mathbb{V}ar}}
\newcommand{\Vbf}{{\bm V}}
\newcommand{\Zbf}{{\bm Z}}
\newcommand{\zbf}{{\bm z}}
\newcommand{\Sigmabf}{\boldsymbol{\Sigma}}

\title{Debiased machine learning for partially linear accelerated failure time models}
\author{Tomoki Okuno, Sijie Zheng, Brendon Chau,\\ Gang Li, Jin J. Zhou and Hua Zhou\thanks{Corresponding author. E-mail: \href{mailto:huazhou@ucla.edu}{huazhou@ucla.edu}}\\[4pt]
{\normalsize\itshape Department of Biostatistics, University of California, Los Angeles}}
\date{}

\begin{document}
\begingroup\setstretch{1.2}\maketitle\endgroup

\begin{abstract}
The Cox model remains the default for survival analysis, but the proportional hazards assumption is often violated and hazard ratios can be difficult to interpret.
Accelerated failure time (AFT) models provide an intuitive time-scale alternative, yet flexible covariate adjustment while preserving valid inference on a target exposure remains challenging.
For the partially linear AFT model under right censoring, a rank-based debiased machine learning (DML) framework has not been developed. The rank-based pairwise moment is not Neyman orthogonal, and standard cross-fitting does not directly apply to U-statistics.
We develop the first such framework by combining an orthogonalized rank-based U-statistic, a censoring-corrected influence function, and block-pairwise cross-fitting, yielding valid inference for a finite-dimensional exposure effect under flexible machine-learning nuisance estimation.
In the primary simulation study, the proposed estimator attains near-nominal coverage under independent and covariate-dependent censoring at rates up to $60\%$, whereas the nonorthogonal plug-in estimator remains biased.
Using All of Us electronic health record data from adults with diabetes and advanced chronic kidney disease, we apply the proposed method to estimate the associations of serum albumin,
estimated glomerular filtration rate, and systolic blood pressure with incident cardiovascular outcomes.

\end{abstract}

\renewcommand{\theequation}{\thesection.\arabic{equation}}

\section{Introduction}\label{sec:Introduction}
The Cox model \citep{cox_regression_1972} remains the default for time-to-event analysis, but its proportional hazards assumption is frequently violated and hazard ratios can be difficult to interpret \citep{hernan_hazards_2010}.
Modern data sources such as electronic health records and biobank studies further introduce high-dimensional covariates with complex effects, making functional-form misspecification a practical concern for conventional regression-based adjustment \citep{harrell_regression_2015}.

Accelerated failure time (AFT) models \citep{kalbfleisch_1980_statistical} avoid the proportional hazards assumption and offer a direct time-scale interpretation. The partially linear AFT (PL-AFT) model further accommodates complex covariate effects: a primary exposure enters linearly as the inferential target, while the remaining covariates are absorbed in a flexible, nonparametric component estimated via machine learning (ML).

Methodologically, we build on the double/debiased machine learning (DML) framework \citep{chernozhukov_doubledebiased_2018} and the partialling-out principle \citep{robinson_root-n-consistent_1988}. In uncensored partially linear models, the resulting moment is Neyman orthogonal \citep{neyman1959optimal}: it is first-order insensitive to nuisance perturbations at the truth. This property, combined with cross-fitting \citep{schick_asymptotically_1986}, delivers valid inference under weak regularity conditions \citep{newey_asymptotic_1994}. Our objective is to extend these guarantees to right-censored outcomes.

For right-censored outcomes, we use a rank-based U-statistic estimator built on pairwise comparisons \citep{halmos_theory_1946, prentice_linear_1978}. Compared with least-squares alternatives \citep[e.g.,][]{buckley_linear_1979}, rank-based methods lose only modest efficiency under (log-)normal errors but gain substantially under skewed or heavy-tailed distributions \citep{chung_tutorial_2013, kim_deep_2024}. However, as the partialling-out pairwise moment lacks Neyman orthogonality under right censoring, naive plug-in ML estimates can introduce first-order bias. In addition, observation-level cross-fitting alone does not decouple pairwise moments. Building on the DML U-statistic theory \citep{escanciano_debiased_2026}, we construct an orthogonal pairwise moment and implement block-pairwise cross-fitting to achieve valid inference.

Classical partially linear survival models rely on low-dimensional smoothing and have not been designed for data-adaptive ML nuisance functions \citep{qin_censored_2001, wang_empirical_2002, orbe_censored_2003, zou_semiparametric_2011, wang_empirical_2013, aydin_modified_2018, ahmed_penalty_2022, yilmaz_modified_2023}. \citet{chen_rank_2005} obtain rank-based inference for a partially linear censored regression model by comparing observations within fine strata of the nonlinear covariate, which is feasible only for a low-dimensional covariate.

Recent orthogonality-based approaches in survival analysis develop Neyman-orthogonal or doubly robust estimators for censored outcomes \citep{cui_estimating_2023, xu_estimating_2024, frauen_orthogonal_2025}.
However, these methods target heterogeneous treatment effects on the survival-probability or restricted-mean-survival-time scale, and do not extend to rank-based U-statistic inference for a finite-dimensional regression coefficient in a semiparametric AFT model. \citet{escanciano_debiased_2026} develop a general DML theory for U-statistics but do not address right-censored outcomes. To our knowledge, this is the first DML framework for rank-based U-statistic inference in survival analysis.

This paper makes three contributions. First, we develop an orthogonalized rank-based U-statistic with a censoring-corrected influence function for the PL-AFT model.
Second, we adopt a block-pairwise cross-fitting scheme that decouples pairwise moments while retaining the sample efficiency of standard cross-fitting. Third, we establish $\sqrt{n}$-consistency and valid asymptotic inference on the primary exposure effect under ML nuisance estimation, confirm finite-sample performance through simulations, and illustrate the method with All of Us data.

The rest of this paper is organized as follows. Section~\ref{sec:methods} introduces the proposed estimator. Section~\ref{sec:inference} develops its asymptotic properties. Section~\ref{sec:simulation} reports simulation results. Section~\ref{sec:app} presents the All of Us data application. Section~\ref{sec:discussion} concludes with a discussion. Technical details and additional results are in the Supplementary Material.

\section{Model and Estimation Method}\label{sec:methods}

\subsection{Notation}
We write $a\wedge b=\min\{a,b\}$, $y(t)=\log t$ for $t>0$, and $f(u-)$ for the left limit of a function $f$ at $u$.
For nonnegative sequences $\{a_n\}$ and $\{b_n\}$, $a_n\lesssim b_n$ or $a_n=O(b_n)$ means $a_n\le c\,b_n$ for some constant $c>0$, and $a_n=o(b_n)$ means $a_n/b_n\to 0$. We write $\overset{p}\to$ and $\overset{d}\to$ for convergence in probability and in distribution. For a random sequence $\{A_n\}$, $A_n=o_p(b_n)$ means $A_n/b_n\overset{p}\to 0$, and $A_n=O_p(b_n)$ means that $A_n/b_n$ is bounded in probability.
For a map $a\mapsto\Psi(a)$, we write $\partial_a\Psi(a)[h]$ for the G\^ateaux derivative in the direction $h$, evaluated at $a$. When $a$ is finite-dimensional, $\partial_a\Psi(a)$ denotes the usual (partial) derivative and $[h]$ is omitted.

\subsection{Problem setup}
We consider the semiparametric PL-AFT model
\begin{equation}
\begin{aligned}
    y(T) = \log(T) &= X\beta_0 + f_0(\Zbf) + \epsilon, \qquad \E[\epsilon] = 0, \\
    X &= m_0(\Zbf) + \nu, \qquad \E[\nu\given \Zbf] = 0,
    \label{eqn:PLTM}
\end{aligned}
\end{equation}
where $T \in\mathbb{R}^+$ is the event time, $X\in \mathbb{R}$ is the exposure of primary interest (binary or continuous), $\beta_0$ is the target regression parameter, $\Zbf\in\mathcal{Z}\subset\mathbb{R}^q$ is a vector of background covariates, and $\epsilon$ and $\nu$ are error terms. The functions $f_0: \mathbb{R}^q\to\mathbb{R}$ and $m_0: \mathbb{R}^q\to\mathbb{R}$ are unknown nuisance components. Here, $f_0$ captures the potentially complex association between $\Zbf$ and the outcome, while $m_0(\Zbf) = \E[X \given \Zbf]$ represents the conditional mean of the exposure. Throughout, the intercept is absorbed into $f_0(\cdot)$.
While $X$ could be generalized to a low-dimensional vector of exposures, we focus on a single interpretable effect. The parameter $\beta_0$ admits a direct time-scale interpretation: a one-unit increase in $X$ multiplies every conditional quantile of $T$ by $\exp(\beta_0)$, holding $\Zbf$ fixed.

In survival analysis, the event time $T$ is often subject to right censoring. Let $C \in \mathbb{R}^+$ denote the censoring time. Instead of $T$ itself, we observe the quadruple $\mathcal{D} = (\tilde y, \Delta, X, \Zbf)$, where $\tilde y = y(T \wedge C)$ and $\Delta = \I\{T \le C\}$. We write $\{\mathcal{D}_i\}_{i=1}^n$ for $n$ independent and identically distributed (i.i.d.) copies of $\mathcal{D}$. Throughout, we often omit the subscript $i$ and write $\mathcal D'$ for an independent copy of $\mathcal D$.

We assume that the full data $(T,C,X,\Zbf)$ are drawn from a population distribution $\mathcal P$.
All expectations $\E[\cdot]$ and probabilities $\prob(\cdot)$ are taken with respect to $\mathcal P$, and we use $x$ and $\zbf$ for realizations of $X$ and $\Zbf$, respectively. For a measurable function $h$ of $\mathcal D$, $\|h\|_2=\{\E[h(\mathcal D)^2]\}^{1/2}$ denotes the $L_2(\mathcal P)$ norm.

We impose the following conditions for identification of the target parameter $\beta_0$.

\begin{ass}%
\label{ass:basic_id}
(a) $\epsilon\perp(X,\Zbf)$.
(b) There exists a constant $c>0$ such that $c\le\E[\nu^2]<\infty$.
\end{ass}
\noindent
Assumption~\ref{ass:basic_id}(a) requires a common error distribution across values of $(X,\Zbf)$ and, with $\E[\epsilon]=0$ in \eqref{eqn:PLTM}, implies $\E[\epsilon\given X,\Zbf]=0$.
Apart from $\E[\nu\given\Zbf]=0$ and (b), the conditional distribution of $\nu$ given $\Zbf$ is unrestricted, so $\nu$ may be heteroscedastic.

\begin{ass}%
\label{ass:cens_id}
(a) $T\perp C\given X,\Zbf$.
(b) For every $(x,\zbf)$ and every $s$ with $\prob(y(T) > s \given X=x, \Zbf=\zbf) > 0$,
$
\prob(y(C) > s \given X=x, \Zbf=\zbf) > 0.
$
(c) $\E\bigl[\,\lvert y(T)\rvert\,\bigr] < \infty$.
\end{ass}
\noindent
Assumption~\ref{ass:cens_id}(a) allows the censoring mechanism to be handled separately from $\beta_0$, (b) ensures that the censoring-corrected influence function of Section~\ref{subsec:orthogonality} is well defined over the support of $y(T)$, and (c) makes the conditional mean $\ell_0$ introduced below well defined.

To estimate $\beta_0$, we adopt the partialling-out principle \citep{robinson_root-n-consistent_1988}. Taking conditional expectations of \eqref{eqn:PLTM} given $\Zbf=\zbf$ gives $\ell_0(\zbf)=m_0(\zbf)\beta_0+f_0(\zbf)$, where $\ell_0(\zbf)=\E[y(T)\given \Zbf=\zbf]$. Estimating $f_0$ would require joint estimation with $\beta_0$. Because $\ell_0$ does not involve $\beta_0$, we estimate $\ell_0$ and $m_0$ flexibly and then estimate $\beta_0$ from
$
    y(T) - \ell_0(\Zbf) = \{X - m_0(\Zbf)\}\beta_0 + \epsilon,
$
which under Assumption~\ref{ass:basic_id}(a) is a semiparametric AFT model with regressor $X-m_0(\Zbf)$ and a common error distribution.

\subsection{Rank-based U-statistic estimating equation}\label{subsec:rank}

The rank-based approach for AFT models is formalized through weighted log-rank estimating equations \citep{prentice_linear_1978}, with large-sample theory developed by \citet{tsiatis_estimating_1990}. Among available weighting schemes, the Gehan weight \citep{gehan_generalized_1965} is preferred because it yields a monotone estimating equation \citep{fygenson_monotone_1994} that is the gradient of a convex loss function \citep{jin_rankbased_2003}.

We extend this framework to the PL-AFT model \eqref{eqn:PLTM}. Define $e_i(\beta,\ell,m)=\tilde y_i-\ell(\Zbf_i)-\{X_i-m(\Zbf_i)\}\beta$.
Let $X_{ij}(m) = \{X_i-m(\Zbf_i)\}-\{X_j-m(\Zbf_j)\}$. The Gehan-weighted rank estimating equation is constructed from the pairwise moment function
\begin{align*}
    g^*(\mathcal D_i,\mathcal D_j;\beta,\ell,m) = \Delta_i\,\I\{e_i(\beta,\ell,m)\le e_j(\beta,\ell,m)\}X_{ij}(m),
\end{align*}
which satisfies $\E[g^*(\mathcal D_i,\mathcal D_j;\beta_0,\ell_0,m_0)]=0$ under Assumptions~\ref{ass:basic_id} and~\ref{ass:cens_id}(a) (Lemma~\ref{lem:gehan-unbiased}). We then define the associated V-statistic estimating function by
$
V_n[g^*(\beta;\ell,m)] = n^{-2} \sum_{i=1}^n\sum_{j=1}^n
g^*(\mathcal D_i,\mathcal D_j;\beta,\ell,m).
$
The estimator for $\beta_0$ is obtained as a solution in $\beta$ to $V_n[g^*(\beta;\ell_0,m_0)]=0$ when the nuisance functions are known (or consistently estimated).
However, the indicator $\I\{e_i(\beta,\ell,m)\le e_j(\beta,\ell,m)\}$ makes $g^*$ discontinuous, which can lead to unstable root-finding and does not permit direct differentiation.
We impose the following regularity condition on pairwise residual differences.

\begin{ass}
\label{ass:rank-M2}
For $i\neq j$, let $D_{ij,0}=e_j(\beta_0,\ell_0,m_0)-e_i(\beta_0,\ell_0,m_0)$ and $\Vbf_k=(X_k,\Zbf_k)$. There exist $L_0<\infty$ and a function $q_{ij}$ such that, almost surely, $\prob(\Delta_i=1,D_{ij,0}\le t\given\Vbf_i,\Vbf_j)=\int_{-\infty}^t q_{ij}(s\given\Vbf_i,\Vbf_j)\,ds$ for all $t$ and $\sup_t q_{ij}(t\given\Vbf_i,\Vbf_j)\le L_0$.
\end{ass}
\noindent
Assumption~\ref{ass:rank-M2} rules out ties between an uncensored residual and any other residual. We use it below to bound the induced-smoothing approximation error and to justify the subsequent differentiation arguments.

We apply induced smoothing \citep{johnson_induced_2009} by replacing the indicator with the smooth approximation
\begin{equation}
\tilde g^*(\mathcal D_i,\mathcal D_j;\beta,\ell,m)
=\Delta_i\,\Phi\!\left(\frac{e_j(\beta,\ell,m)-e_i(\beta,\ell,m)}{\Gamma_n^{1/2}\lvert X_{ij}(m)\rvert}\right)X_{ij}(m),
\label{eqn:rank-smooth-moment}
\end{equation}
where $\Phi(\cdot)$ is the standard normal distribution function and the smoothing parameter $\Gamma_n>0$ satisfies $\Gamma_n=O(n^{-1})$.
When $X_{ij}(m)=0$, functions of this ratio are replaced by their limits as $\Gamma_n^{1/2}\lvert X_{ij}(m)\rvert\to0$, so that $\Phi(\cdot)$ becomes $\I\{e_j(\beta,\ell,m)>e_i(\beta,\ell,m)\}$ and $\tilde g^*=0$.
Lemma~\ref{lem:induced-smoothing} justifies this replacement at the population level: the smoothed and nonsmoothed moments differ by a vanishing approximation error, so they target the same identifying condition asymptotically.

\begin{lemma}[Induced-smoothing approximation]
\label{lem:induced-smoothing}
Under Assumptions~\ref{ass:basic_id}(b) and~\ref{ass:rank-M2}, for any fixed $(\beta,\ell,m)$ with $\|m-m_0\|_2<\infty$,
$
\big|\E\!\left[\tilde g^*(\mathcal D,\mathcal D';\beta,\ell,m)\right]
-\E\!\left[g^*(\mathcal D,\mathcal D';\beta,\ell,m)\right]\big|
\;\lesssim\;\Gamma_n^{1/2}.
$
With $\Gamma_n=O(n^{-1})$, the difference is $O(n^{-1/2})$.
\end{lemma}

The rank-based estimator is often written in terms of the (smoothed) V-statistic estimating equation.
To leverage the DML U-statistic theory \citep{escanciano_debiased_2026}, however, we work with a U-statistic representation based on the symmetrized smoothed moment function
\begin{align}
\tilde g(\mathcal D_i,\mathcal D_j;\beta,\ell,m)
=\frac12\left\{\tilde g^*(\mathcal D_i,\mathcal D_j;\beta,\ell,m)+\tilde g^*(\mathcal D_j,\mathcal D_i;\beta,\ell,m)\right\},
\label{eqn:rank-symmetric-moment}
\end{align}
which is symmetric in $(\mathcal D_i,\mathcal D_j)$ and facilitates orthogonality arguments (Section~\ref{subsec:orthogonality}). The corresponding U-statistic estimating function is
\begin{align}
U_n[\tilde g(\beta;\ell,m)]
=\frac{2}{n(n-1)}\sum_{1\le i<j\le n}\tilde g(\mathcal D_i,\mathcal D_j;\beta,\ell,m).
\label{eqn:rank-U-statistics}
\end{align}
Since the diagonal terms vanish (i.e., $\tilde g(\mathcal D_i,\mathcal D_i;\cdot)=0$), $V_n[\tilde g(\beta;\ell,m)]$ is proportional to $U_n[\tilde g(\beta;\ell,m)]$ and therefore yields the same root in $\beta$.

When $(\ell_0,m_0)$ are replaced by ML estimators, the plug-in moment $U_n[\tilde g(\beta;\hat\ell,\hat m)]$ is not Neyman orthogonal: its pathwise derivative with respect to $(\ell,m)$ at $(\ell_0,m_0)$ is generally nonzero, so first-step estimation errors can bias inference (see Section~\ref{subsec:orthogonality}).

\begin{remark}
\label{rmk:LS-alternative}
An alternative outcome-imputation approach would first transform the censored outcome into a pseudo-outcome that preserves $\E[y(T)\given X,\Zbf]$, and then apply the uncensored partialling-out moment. Such transformations include inverse or log-censoring weighted outcomes \citep{zheng_class_1987}, conditional imputation of the latent outcome \citep{buckley_linear_1979}, and doubly robust combinations of weighting and imputation \citep{rubin_doubly_2007}. These approaches target a least-squares-type moment rather than a rank-based moment and require additional modeling of the censoring or tail outcome distribution. We focus instead on the rank-based formulation, which exploits observed residual orderings and avoids direct reconstruction of $y(T)$, but requires an explicit orthogonality correction for ML nuisance estimation. Such a transformation enters our procedure only through the pseudo-outcome regression for $\ell_0$ (Algorithm~\ref{alg:rank}) and the orthogonality correction (Section~\ref{subsec:orthogonality}), while $\beta_0$ is identified by the rank-based moment.
\end{remark}

\subsection{Constructing an orthogonal moment}\label{subsec:orthogonality}
We next orthogonalize the smoothed rank moment \eqref{eqn:rank-symmetric-moment} with respect to the first-step estimation of $\ell_0$ and $m_0$. The construction adds a projected sensitivity correction to $\tilde g$, following the DML U-statistic approach of \citet{escanciano_debiased_2026}.

With Assumption~\ref{ass:rank-M2} in place, define pairwise sensitivity weights for $\tilde g$ at the truth. Let $E_{ij,0}=D_{ij,0}\{\Gamma_n^{1/2}\lvert X_{ij}(m_0)\rvert\}^{-1}$ and $X_{ij,0}=X_{ij}(m_0)$ for brevity, and let $\phi(\cdot)$ denote the standard normal density function. Define
\begin{equation}
\begin{aligned}
W_{ij,\ell}
&=\frac{\Delta_i+\Delta_j}{2}\,
  \phi(E_{ij,0})\,
  \frac{\mathrm{sgn}(X_{ij,0})}{\Gamma_n^{1/2}},\\[4pt]
W_{ij,m}
&=\frac{\Delta_i+\Delta_j}{2}
  \left[\phi(E_{ij,0})
  \left\{E_{ij,0}
  -\frac{\beta_0\,\mathrm{sgn}(X_{ij,0})}{\Gamma_n^{1/2}}\right\}
  -\Phi(E_{ij,0})\right]
  +\frac{\Delta_j}{2}.
\end{aligned}
\label{eqn:pairwise-weights}
\end{equation}
These weights are antisymmetric ($W_{ji,\bullet}=-W_{ij,\bullet}$, $\bullet\in\{\ell,m\}$) and represent the G\^ateaux derivatives of $\E[\tilde g]$ at the truth (Section~\ref{app:proof-rank-orthogonal}). Their projections are $\alpha_{\bullet,0}(\zbf)=2\,\E[W_{ij,\bullet}\given\Zbf_i=\zbf]$, where the factor $2$ follows from antisymmetry and the identical distribution of $\mathcal D_i$ and $\mathcal D_j$.
The adjustment term is then given by
\begin{align}
\gamma(\mathcal{D}_i, \mathcal{D}_j;\, \eta, \alpha_\ell, \alpha_m)
&= \frac{1}{2}\sum_{k\in\{i,j\}}\Big[
  \alpha_\ell(\Zbf_k)\,\varphi^\star_k(\ell, G, S^T)
  + \alpha_m(\Zbf_k)\{X_k - m(\Zbf_k)\}\Big],
\label{eq:gamma-ij}
\end{align}
where $\eta = (\ell, m, G, S^T)$ has true value $\eta_0=(\ell_0,m_0,G_0,S^T_0)$ with $G_0(s\given x,\zbf)=\prob(y(C)>s\given X=x,\Zbf=\zbf)$ and $S^T_0(s\given x,\zbf)=\prob(y(T)>s\given X=x,\Zbf=\zbf)$. The exposure residual $X-m(\Zbf)$ is observed, whereas the outcome residual $y(T)-\ell(\Zbf)$ is not. We therefore replace the latter with a censoring-corrected influence function $\varphi^\star$, constructed below.

Define the censoring residual process under a generic censoring law $G$ by $M^C_G(t)=\I\{\tilde y\le t,\,\Delta\!=\!0\}-\int_{-\infty}^{t}\I\{\tilde y\ge s\}\,d\Lambda^C_G(s\given X,\Zbf)$, where $d\Lambda^C_G(s\given X,\Zbf)=-dG(s\given X,\Zbf)/G(s-\given X,\Zbf)$ is the cumulative-hazard increment of $G$, equal to $-d\log G(s\given X,\Zbf)$ when $G$ is continuous. Following \citet{overgaard_comparison_2025}, the influence function has the form
\begin{equation}
\varphi^\star(\ell, G, S^T)
= \xi^\star(\ell, G)
  +\int_{-\infty}^{\infty}
    \frac{Q^\star(u,X,\Zbf;\, \ell, S^T)}{G(u\given X,\Zbf)}\,dM^C_G(u),
\label{eq:phi-general}
\end{equation}
where $\xi^\star$ satisfies $\E[\xi^\star(\ell_0,G_0)\given\Zbf]=0$ and $Q^\star$ is chosen so that $\partial_G\E[\varphi^\star(\ell_0,G_0,S^T_0)\given\Zbf][H]=0$ for every admissible direction $H$. Section~\ref{app:G-orthogonality} specifies the admissible directions, gives the cancellation condition that determines $Q^\star$, and verifies it for the construction below.

We take $\xi^\star$ to be the inverse-probability-of-censoring-weighted (IPCW) residual, and $Q^\star$ is then determined by the cancellation condition,
\begin{align}
\xi^{I}(\ell,G)
= \frac{\Delta\bigl\{\tilde y - \ell(\Zbf)\bigr\}}{G(\tilde y\given X,\Zbf)}, \quad
Q^{I}(u,x,\zbf;\,\ell, S^T)
= u + \frac{\int_{u}^{\infty} S^T(s\given x,\zbf)\,ds}{S^T(u\given x,\zbf)} - \ell(\zbf).
\label{eq:ipcw-residual}
\end{align}
At the truth, $Q^{I}(u,x,\zbf;\ell_0,S^T_0)=\E[y(T)-\ell_0(\Zbf)\given y(T)\ge u,X=x,\Zbf=\zbf]$ is the mean residual among subjects still at risk at $u$. Since $\ell$ enters additively, $\varphi^\star(\ell,G,S^T)=\rho^\star(G,S^T)-\ell(\Zbf)$, where $\rho^\star$ does not depend on $\ell$ and $\rho_0=\rho^\star(G_0,S^T_0)$ satisfies $\E[\rho_0\given\Zbf]=\ell_0(\Zbf)$.

At the truth, $\varphi_0=\varphi^\star(\ell_0,G_0,S^T_0)$ satisfies $\E[\varphi_0\given\Zbf]=0$, since $\xi^{I}$ has $\Zbf$-conditional mean zero and the martingale integral has mean zero by the predictability of its integrand.
Therefore, $\E[\gamma(\mathcal D_i,\mathcal D_j;\eta_0,\alpha_{\ell,0},\alpha_{m,0})]=0$.
In computation, $\varphi^\star$ is replaced by a horizon-specific version $\varphi^\star_U=\rho^\star_U(G,S^T)-\ell(\Zbf)$ with a data-dependent horizon $U\to\infty$, constructed to preserve orthogonality at a finite horizon (Sections~\ref{app:nuisance_estimation} and~\ref{app:G-orthogonality}).
We define the orthogonalized smoothed moment by adding the correction term~\eqref{eq:gamma-ij}:
\begin{align}
\tilde\psi(\mathcal D_i,\mathcal D_j;\beta,\eta,\alpha_\ell,\alpha_m)
=\tilde g(\mathcal D_i,\mathcal D_j;\beta,\ell,m)
+\gamma(\mathcal D_i,\mathcal D_j;\eta,\alpha_\ell,\alpha_m).
\label{eqn:rank-orthogonal-moment}
\end{align}

\begin{prop}[Neyman orthogonality]
\label{prop:rank-orthogonal}
Let $\omega=(\eta,\alpha_\ell,\alpha_m)$ and $\omega_0=(\eta_0,\alpha_{\ell,0},\alpha_{m,0})$. Suppose that Assumptions~\ref{ass:basic_id}--\ref{ass:rank-M2} hold, that $y(T)$ has a conditional density given $(X,\Zbf)$, and that $\E[\varphi_0^2]<\infty$. Then, for every admissible direction $h$,
$
\partial_\omega\E[\tilde\psi(\mathcal D,\mathcal D';\beta_0,\omega_0)][h]
=\frac{d}{dr}\E\big[\tilde\psi(\mathcal D,\mathcal D';\beta_0,\omega_0+rh)\big]\Big|_{r=0}=0,
$
where $h$ is admissible if its components for $(\ell,m,\alpha_\ell,\alpha_m)$ are square-integrable functions of $\Zbf$, its components for $(G,S^T)$ belong to the classes in Section~\ref{app:G-orthogonality}, and at most one of its $G$- and $S^T$-components is nonzero.
\end{prop}

The functions $\alpha_{\ell,0}$ and $\alpha_{m,0}$ are the Riesz representers of the linear functionals induced by the G\^ateaux derivatives of $\E[\tilde g]$ with respect to $\ell$ and $m$ \citep{chernozhukov_debiased_2022}. In practice, we estimate them by regressing pseudo-outcomes constructed from $W_{ij,\bullet}$ on $\Zbf$.

\subsection{Block-pairwise cross-fitting}\label{subsec:cross-fitting}
The orthogonal moment $\tilde\psi$ removes first-order sensitivity to nuisance perturbations, but using the same observations for nuisance fitting and moment evaluation can still induce overfitting bias.
We use block-pairwise cross-fitting \citep[Figures~1 and 9]{escanciano_debiased_2026} to address this.

Let $K\ge 2$ be a fixed integer. We partition $\{1,\ldots,n\}$ into $K$ groups of nearly equal size and split each group into two nearly equal halves. The pairs within groups form $K$ triangles, and the pairs between halves of different groups form $2K(K-1)$ rectangles, yielding $L=K(2K-1)$ disjoint blocks $\{\mathcal B_1,\ldots,\mathcal B_L\}$ of the unordered index pairs $\{(i,j)\in\{1,\ldots,n\}^2:i<j\}$.
For each block $l$, we fit nuisance estimators using only observations that do not appear in any pair in $\mathcal B_l$, and denote the resulting fits by $\hat\eta^{(l)}=(\hat\ell^{(l)}, \hat m^{(l)}, \hat{G}^{(l)}, \hat S^{T,(l)})$, $\hat\alpha_\ell^{(l)}$, and $\hat\alpha_m^{(l)}$.

To estimate $\alpha_{\ell,0}$ and $\alpha_{m,0}$ within each block $l$, we construct observation-level pseudo-outcomes from the pairwise weights in~\eqref{eqn:pairwise-weights}.
Let $\mathcal I_l=\bigcup_{(i,j)\in\mathcal B_l}\{i,j\}$ and
$\mathcal I_{-l}=\{1,\ldots,n\}\setminus \mathcal I_l$. For each $i\in\mathcal I_{-l}$, define the partner set
$\mathcal J_{-l}(i)=\mathcal I_{-l}\setminus\{i\}$, and construct the pseudo-outcomes
\begin{align}
\hat T^{\ell,(l)}_i
=\frac{2}{|\mathcal J_{-l}(i)|}\sum_{j\in\mathcal J_{-l}(i)} \hat W^{(l)}_{ij,\ell},
\qquad
\hat T^{m,(l)}_i
=\frac{2}{|\mathcal J_{-l}(i)|}\sum_{j\in\mathcal J_{-l}(i)} \hat W^{(l)}_{ij,m},
\label{eqn:pseudo-outcome}
\end{align}
where $\hat W^{(l)}_{ij,\bullet}$ is evaluated with a pilot $\hat\beta^{(0,l)}$ solving the unaugmented estimating equation~\eqref{eqn:rank-U-statistics} over pairs in $\mathcal I_{-l}$ and with predictions of $(\ell_0,m_0)$ cross-fitted within $\mathcal I_{-l}$.
We Rao--Blackwellize these pseudo-outcomes over the fitted laws of $\epsilon_i$ and $y(C_i)$ given $(X_i,\Zbf_i)$, holding the partners fixed, and denote the results by $\bar T^{\bullet,(l)}_i$ (Section~\ref{app:nuisance_estimation}). At the truth, this preserves $\E[\bar T^{\bullet,(l)}_i\given\Zbf_i]=\alpha_{\bullet,0}(\Zbf_i)$ and does not increase the conditional variance.
Regressing $(\bar T^{\ell,(l)}_i, \bar T^{m,(l)}_i)$ on $\Zbf_i$ over $i\in\mathcal I_{-l}$ yields $(\hat\alpha_\ell^{(l)}, \hat\alpha_m^{(l)})$.

The cross-fitted estimating equation is
\begin{align}
U_n^{\rm cf}[\tilde\psi(\beta)]
=\frac{2}{n(n-1)}\sum_{l=1}^{L}\ \sum_{(i,j)\in\mathcal B_l}
\tilde\psi\!\left(\mathcal D_i,\mathcal D_j;\beta,\hat\eta^{(l)}, \hat\alpha_\ell^{(l)}, \hat\alpha_m^{(l)}\right),
\label{eqn:cross-fitted-U-statistics}
\end{align}
and the cross-fitted U-estimator $\hat\beta$ is defined as a solution to $U_n^{\rm cf}[\tilde\psi(\beta)]=0$.

The estimation procedure is summarized in Algorithm~\ref{alg:rank}. The pair blocks are built from a random partition of the indices $\{1,\ldots,n\}$ that does not depend on the data. Every fitted quantity used for block $l$, including $\hat\beta^{(0,l)}$, depends on the data only through $\mathcal I_{-l}$.

\begin{algorithm}[htbp]
\label{alg:rank}
\KwIn{dataset $\{\mathcal D_i\}_{i=1}^n$; pair blocks $\{\mathcal B_l\}_{l=1}^L$ from a random partition of $\{1,\ldots,n\}$.}

Define $\mathcal I_l \gets \bigcup_{(i,j)\in\mathcal B_l}\{i,j\}$ and
$\mathcal I_{-l}\gets \{1,\ldots,n\}\setminus\mathcal I_l$ for all $l=1,\ldots,L$.\\
Initialize $U_n^{\rm cf}[\tilde\psi(\beta)]\gets 0$.\\
\For{$l=1$ \KwTo $L$}{
  Fit $\hat m^{(l)}$, $\hat G^{(l)}$ and $\hat S^{T,(l)}$ on $\mathcal I_{-l}$.\\
  Fit $\hat\ell^{(l)}$ by regressing $\hat\rho_i=\rho^\star_{U}(\hat G,\hat S^T)$ on $\Zbf_i$ over $i\in\mathcal I_{-l}$, with $(\hat G,\hat S^T,U)$ cross-fitted within $\mathcal I_{-l}$ (Section~\ref{app:nuisance_estimation}).\\
  Obtain the pilot $\hat\beta^{(0,l)}$ as a solution to $\sum_{i<j,\;i,j\in\mathcal I_{-l}}\tilde g(\mathcal D_i,\mathcal D_j;\beta,\hat\ell,\hat m)=0$, with $(\hat\ell,\hat m)$ cross-fitted within $\mathcal I_{-l}$.\\
  Compute $\hat\varphi_k^{(l)}=\varphi^\star_{U_{n,l}}(\hat\ell^{(l)},\hat G^{(l)},\hat S^{T,(l)})$ for $k\in\mathcal I_l$, with the horizon $U_{n,l}$ computed from $\mathcal I_{-l}$.\\
  Construct the Rao--Blackwellized versions $\{\bar T^{\ell,(l)}_i,\bar T^{m,(l)}_i:i\in\mathcal I_{-l}\}$ of the pseudo-outcomes in~\eqref{eqn:pseudo-outcome} (Section~\ref{app:nuisance_estimation}).\\
  Fit $\hat\alpha_\ell^{(l)}$ and $\hat\alpha_m^{(l)}$ by regressing $\bar T^{\ell,(l)}_i$ and $\bar T^{m,(l)}_i$ on $\Zbf_i$ over $i\in\mathcal I_{-l}$.\\
  $U_n^{\rm cf}[\tilde\psi(\beta)]\gets U_n^{\rm cf}[\tilde\psi(\beta)] + \frac{2}{n(n-1)}\sum_{(i,j)\in\mathcal B_l}
  \tilde\psi(\mathcal D_i,\mathcal D_j;\beta,\hat\eta^{(l)},\hat\alpha_\ell^{(l)},\hat\alpha_m^{(l)})$.\\
}
\Return{$\hat\beta$ as a solution to $U_n^{\rm cf}[\tilde\psi(\hat\beta)]=0$.}
\caption{Rank-based DML U-estimator for the PL-AFT model.}
\end{algorithm}

\begin{remark}
The pilot $\hat\beta^{(0,l)}$ is used only to construct the Rao--Blackwellized pseudo-outcomes.
Any consistent estimator of $\beta_0$ that depends on the data only through $\mathcal I_{-l}$ suffices.
\end{remark}

\begin{remark}
The block-pairwise cross-fitting scheme partitions the index-pair set into $L=K(2K-1)$ disjoint blocks. Only $K$ of these blocks are diagonal (triangle) blocks, which correspond to the $K$ folds in ordinary $K$-fold cross-fitting. Compared to ordinary $K$-fold cross-fitting, the block-pairwise construction requires $L/K=2K-1\ge 3$ times as many nuisance refits.
\end{remark}

\section{Asymptotic Properties and Variance Estimation}\label{sec:inference}

We establish $\sqrt{n}$-consistency and asymptotic normality of the rank-based estimator $\hat\beta$ based on the orthogonal moment~\eqref{eqn:rank-orthogonal-moment}.
We impose the following regularity conditions.

\begin{ass}%
\label{ass:smoothing-rate}
The function $t\mapsto q_{ij}(t\given \Vbf_i,\Vbf_j)$ in Assumption~\ref{ass:rank-M2} is Lipschitz with constant $L_1<\infty$, almost surely.
\end{ass}
\noindent
Assumption~\ref{ass:smoothing-rate} strengthens Assumption~\ref{ass:rank-M2} for two purposes. It sharpens the induced-smoothing bias at the truth (Lemma~\ref{lem:induced-smoothing-sharp}) and controls the second-order remainder in the cross-fitting expansion (Lemma~\ref{lem:first-step}).

\begin{lemma}[Sharpened induced-smoothing bias]
\label{lem:induced-smoothing-sharp}
Under Assumptions~\ref{ass:basic_id}(b), \ref{ass:rank-M2} and~\ref{ass:smoothing-rate}, at the truth $(\beta_0,\ell_0,m_0)$ and as $\Gamma_n\to0$,
$
\big|\E\!\left[\tilde g^*(\mathcal D,\mathcal D';\beta_0,\ell_0,m_0)\right]
-\E\!\left[g^*(\mathcal D,\mathcal D';\beta_0,\ell_0,m_0)\right]\big|
=o(\Gamma_n^{1/2}).
$
\end{lemma}
\noindent
With $\Gamma_n=O(n^{-1})$, Lemma~\ref{lem:induced-smoothing-sharp}, Lemma~\ref{lem:gehan-unbiased} and $\E[\gamma(\mathcal D,\mathcal D';\omega_0)]=0$ give $\E[\tilde\psi(\mathcal D,\mathcal D';\beta_0,\omega_0)]=o(n^{-1/2})$, so the smoothing does not affect $\sqrt n$ inference.

\begin{ass}%
\label{ass:score-regularity}
(a) There exists a measurable envelope $b$, not depending on $n$,\ such that $|\tilde\psi(\mathcal D_i,\mathcal D_j;\beta_0,\omega_0)|\le b$ and $\E[b^2]<\infty$.
(b) $A=\E[X_{ij}(m_0)^2\,q_{ij}(0\given\Vbf_i,\Vbf_j)]$ is positive.
(c) The limit $V=\lim_{n\to\infty}4\,\V[ \E\{ \tilde\psi(\mathcal D_i,\mathcal D_j;\beta_0,\omega_0) \given \mathcal D_i \}]$, which exists by Section~\ref{app:proof-asymp-normal}, is positive.
\end{ass}

\begin{ass}%
\label{ass:rates-L2}
For block-specific nuisance estimates, all $L_2(\mathcal P)$ norms and expectations below are taken over an independent evaluation draw, conditional on the corresponding training sample and any auxiliary randomness used to construct them.
Let $\hat\eta^{(l)}=(\hat\ell^{(l)},\hat m^{(l)},\hat G^{(l)},\hat S^{T,(l)})$ and $(\hat\alpha_\ell^{(l)},\hat\alpha_m^{(l)})$ denote the out-of-block nuisance estimators in~\eqref{eqn:cross-fitted-U-statistics}, and let $\hat\varphi^{(l)}=\varphi^\star_{U_{n,l}}(\hat\ell^{(l)},\hat G^{(l)},\hat S^{T,(l)})$ denote the estimated censoring-corrected influence function, with the horizon $U_{n,l}$ computed from the training sample of block $l$.
Define block-maximal $L_2(\mathcal P)$ rates
$r_n^m = \max_{l}\|\hat m^{(l)}-m_0\|_{2}$,
$r_n^\varphi = \max_{l}\|\hat\varphi^{(l)}-\varphi_0\|_{2}$,
$r_n^{\alpha_\ell} = \max_{l}\|\hat\alpha_\ell^{(l)}-\alpha_{\ell,0}\|_{2}$,
$r_n^{\alpha_m} = \max_{l}\|\hat\alpha_m^{(l)}-\alpha_{m,0}\|_{2}$, and
$r_n^\chi=\max_l\|\hat\chi^{(l)}-\chi_0\|_2$, where $\chi_0=\alpha_{\ell,0}(\Zbf)\varphi_0+\alpha_{m,0}(\Zbf)\{X-m_0(\Zbf)\}$ is the summand of $\gamma$ in~\eqref{eq:gamma-ij} at the truth and $\hat\chi^{(l)}=\hat\alpha_\ell^{(l)}(\Zbf)\hat\varphi^{(l)}+\hat\alpha_m^{(l)}(\Zbf)\{X-\hat m^{(l)}(\Zbf)\}$.
Each is $o_p(1)$ and
(i) $\max_l\|\hat\ell^{(l)}-\ell_0\|_{2,w}=o_p(n^{-1/4})$ and $\max_l\|\hat m^{(l)}-m_0\|_{2,w}=o_p(n^{-1/4})$, where $\|h\|_{2,w}^2=\E\{w(\Zbf)h(\mathcal D)^2\}$ with $w(\Zbf)=1+\E[\lvert\nu\rvert\given\Zbf]$,
(ii) $r_n^{\alpha_\ell}\,r_n^\varphi = o_p(n^{-1/2})$,
(iii) $r_n^{\alpha_m}\,r_n^m = o_p(n^{-1/2})$, and
(iv) $\max_{l}\bigl|\E[\alpha_{\ell,0}(\Zbf)\{\hat\rho^{(l)}-\rho_0\}]\bigr| = o_p(n^{-1/2})$,
where $\hat\rho^{(l)}=\rho^\star_{U_{n,l}}(\hat G^{(l)},\hat S^{T,(l)})$.
\end{ass}
\noindent
Condition~(i) controls the quadratic remainder of the base rank moment $\tilde g$, which is nonlinear in both $\ell$ and $m$, while conditions~(ii)--(iv) control the adjustment remainder. Conditions~(ii) and (iii) permit product-rate trade-offs between nuisance estimators, analogous to rate double robustness \citep{smucler_unifying_2019}.
Sufficient conditions for the $\ell$-part of condition~(i) and for condition~(iv) are given in Section~\ref{app:G-orthogonality}.
Such high-level conditions are standard in DML \citep{chernozhukov_doubledebiased_2018, chernozhukov_debiased_2022}, although their validity for a particular learner depends on the function class, tuning, and regularity conditions.

\begin{lemma}
\label{lem:first-step}
Under the conditions of Proposition~\ref{prop:rank-orthogonal} and Assumptions~\ref{ass:smoothing-rate}--\ref{ass:rates-L2},
$
U_n^{\rm cf}[\tilde\psi(\beta_0)]
=U_n[\tilde\psi(\beta_0;\omega_0)]+o_p(n^{-1/2}).
$
\end{lemma}
\noindent
Neyman orthogonality (Proposition~\ref{prop:rank-orthogonal}) eliminates the first-order nuisance effect; the product-rate conditions (Assumption~\ref{ass:rates-L2}) control the second-order remainder; cross-fitting removes overfitting bias. The oracle U-statistic on the right-hand side admits a Hoeffding decomposition \citep{hoeffding_class_1948} whose first-order projection $\E[\tilde\psi(\mathcal D_i,\mathcal D_j;\beta_0,\omega_0)\given\mathcal D_i]$ drives the asymptotic distribution by the central limit theorem.

\begin{prop}[Asymptotic normality]
\label{prop:asymp-normal}
Under the conditions of Proposition~\ref{prop:rank-orthogonal} and Assumptions~\ref{ass:smoothing-rate}--\ref{ass:rates-L2}, with $\Gamma_n=O(n^{-1})$ and $\liminf_n n\Gamma_n>0$,
$
\sqrt n(\hat\beta-\beta_0) \overset{d}\to N\!\left(0,\; V / A^2\right),
$
where $A$ and $V$ are defined in Assumption~\ref{ass:score-regularity}(b)--(c).
\end{prop}

Because $\gamma$ does not depend on $\beta$, $A$ is the limit of the Jacobian of the base rank moment~\eqref{eqn:rank-symmetric-moment} (Lemma~\ref{lem:jacobian}). It is estimated by the derivative of the cross-fitted estimating function at $\hat\beta$,
$
\hat A
=\frac{2}{n(n-1)}\sum_{l=1}^{L}\sum_{(i,j)\in\mathcal B_l}
\frac{\Delta_i+\Delta_j}{2}\,
\phi(\hat E^{(l)}_{ij})\,
\frac{\lvert X_{ij}(\hat m^{(l)})\rvert}{\Gamma_n^{1/2}},
$
where $\hat E^{(l)}_{ij}=\{e_j(\hat\beta,\hat\ell^{(l)},\hat m^{(l)})-e_i(\hat\beta,\hat\ell^{(l)},\hat m^{(l)})\}/\{\Gamma_n^{1/2}\lvert X_{ij}(\hat m^{(l)})\rvert\}$.
For $V$, the Hoeffding projection yields
$
\hat V =
\frac{4}{n}\sum_{i=1}^n
\left\{
\frac{1}{n-1}\sum_{j\neq i}
\hat\psi_{ij}
\right\}^2,
$
where $\hat\psi_{ij}=\hat\psi_{ji}=\tilde\psi(\mathcal D_i,\mathcal D_j;\hat\beta,\hat\eta^{(l)},\hat\alpha_\ell^{(l)},\hat\alpha_m^{(l)})$ for $(i,j)\in\mathcal B_l$, with $\hat\alpha_\ell^{(l)}$ and $\hat\alpha_m^{(l)}$ refitted from the Rao--Blackwellized pseudo-outcomes reconstructed at $\hat\beta$ rather than at the block-specific pilots, thereby evaluating the correction term in $\hat V$ at the final estimate, as in \citet{escanciano_debiased_2026}.
The standard error is $\mathrm{SE}(\hat\beta)=\sqrt{n^{-1}\hat V/\hat A^2}$, and a $95\%$ confidence interval is $\hat\beta\pm 1.96\,\mathrm{SE}(\hat\beta)$.
Under the conditions of Proposition~\ref{prop:asymp-normal}, $\hat A\overset{p}\to A$, and $\hat V\overset{p}\to V$ if, in addition, the correction term with the refitted $\hat\alpha_\ell^{(l)}$ and $\hat\alpha_m^{(l)}$ is $L_2$-consistent uniformly over $n^{-1/2}$-neighborhoods of $\beta_0$ (Lemma~\ref{lem:variance}).

\section{Simulation Study}\label{sec:simulation}
This section investigates the finite-sample performance of the proposed estimator. Following \citet{brown_standard_2005}, we set $\Gamma_n=1/n$.

\subsection{Setup}\label{subsec:setup}

We generate data from the PL-AFT model~\eqref{eqn:PLTM} with $n=1000$, $q=50$ covariates and $\beta_0=1$.
Covariates $\Zbf=(Z_{(1)},\ldots,Z_{(50)})$ are drawn from a Gaussian copula with uniform marginals and then standardized, where the latent Gaussian variables have first-order autoregressive correlation 0.5 within non-overlapping blocks of size~10.
The nuisance functions are
$m_0(\Zbf)=0.8\,Z_{(1)}+0.5\,\sin(0.5\pi Z_{(2)})+0.3\,Z_{(3)}+\I\{Z_{(3)}>0\}$ and
$f_0(\Zbf)=0.4\log(1+Z_{(4)}^2)+0.4\cdot\I\{Z_{(3)}>0\}+0.1\cdot\I\{Z_{(5)}>0\}+0.3\,\sin(2\pi Z_{(1)})-0.175\,Z_{(1)}$,
centered and scaled within each sample.
The functions $m_0$ and $f_0$ share $Z_{(1)}$ and $Z_{(3)}$, including the threshold feature $\I\{Z_{(3)}>0\}$, with nonlinear signal embedded in 5 of 50 covariates.
The errors are $\epsilon,\nu\sim\mathcal N(0,1)$, mutually independent and independent of $\Zbf$.
Censoring times follow $\log C=\mu_C+\sigma_C\zeta$ (independent censoring) or $\log C=\mu_C+0.5X-0.5Z_{(1)}+\sigma_C\zeta$ (covariate-dependent censoring), where $\zeta$ follows the standard maximum-type Gumbel distribution independently of $(T,X,\Zbf)$, $\sigma_C=1.5\sqrt6\,\mathrm{sd}\{y(T)\}/\pi$ under independent and $\sigma_C=\sqrt6\,\mathrm{sd}\{y(T)\}/\pi$ under covariate-dependent censoring, and $\mu_C$ is calibrated to the target censoring rate~(CR).
We examine CR $\in\{20\%,40\%,60\%\}$ under both censoring mechanisms, with $500$ replications per setting.
We estimate $m_0$ and $\ell_0$ by XGBoost regression \citep{chen_xgboost_2016} with fixed hyperparameters and the number of boosting rounds chosen by cross-validation, with $\ell_0$ estimated from $\hat\rho_i$ in Algorithm~\ref{alg:rank}, and $\alpha_{\ell,0}$ and $\alpha_{m,0}$ by a generalized additive model in $(\hat\ell,\hat m)$ followed by XGBoost on $\Zbf$.
$S^T(\cdot\given X,\Zbf)$ is estimated by survival forests \citep{athey_generalized_2019}, and $G$ by the Kaplan--Meier estimator under independent censoring and by survival forests under covariate-dependent censoring.
We use $K=3$ folds for block-pairwise cross-fitting ($L=15$ blocks).
Further details, the same settings with $n=500$ and sensitivity scenarios involving exponential and uniform censoring, heteroscedastic $\nu$ and Laplace $\epsilon$ are provided in Section~\ref{app:sim_details}.

We compare five estimators: an oracle estimator solving~\eqref{eqn:rank-U-statistics} with the true $\ell_0$ and $m_0$; a conventional parametric benchmark obtained by maximum likelihood estimation under a log-normal AFT model with $X$ and all 50 covariates entered linearly (Linear AFT); the plug-in estimator solving~\eqref{eqn:rank-U-statistics} with ML nuisance estimates but without cross-fitting; the cross-fitted plug-in (Plug-in CF), which solves~\eqref{eqn:cross-fitted-U-statistics} with $\hat\alpha_\ell^{(l)}=\hat\alpha_m^{(l)}=0$ and thus separates the effect of cross-fitting from that of the orthogonal correction;
and the proposed orthogonalized estimator with block-pairwise cross-fitting (Section~\ref{subsec:cross-fitting}), denoted Orth.

\subsection{Results}\label{subsec:results}
\begin{figure}[htbp]
    \centering
    \includegraphics[width = 0.72\textwidth]{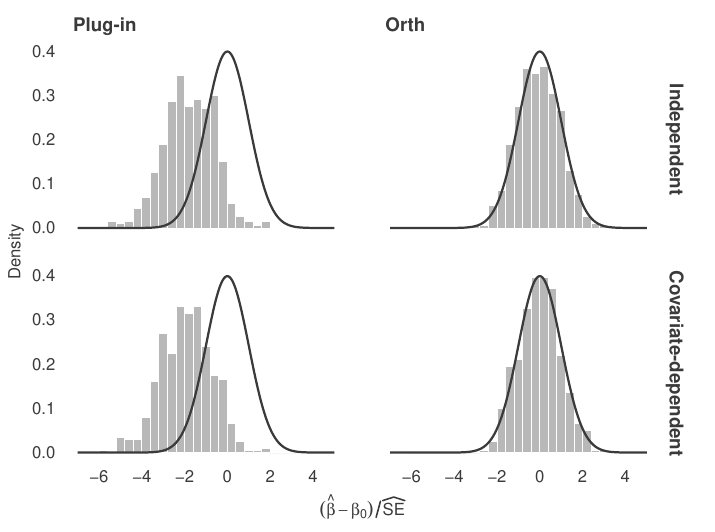}
    \caption{Histograms of $(\hat{\beta}-\beta_0)/\widehat{\mathrm{SE}}$ for the plug-in and proposed estimators, with the $\mathcal{N}(0,1)$ density ($n=1000$, 40\% censoring rate, 500 replications).}
    \label{fig:z_density}
\end{figure}

Figure~\ref{fig:z_density} shows the standardized estimates $(\hat\beta-\beta_0)/\widehat{\mathrm{SE}}$ at CR $40\%$. Under both censoring mechanisms, those of the plug-in estimator are centered near $-2$, whereas those of Orth closely track the $\mathcal{N}(0,1)$ density.
Table~\ref{tab:sim_results} shows the same pattern in all six scenarios.
Linear AFT, which has the correct error distribution but adjusts for $\Zbf$ linearly, has an absolute bias of 0.87 to 1.11 times its SD, with coverage between 0.776 and 0.844. The absolute bias of the plug-in estimator is 1.1 to 1.8 times its SD, with coverage between 0.412 and 0.714. Cross-fitting lowers this bias, but Plug-in CF retains an absolute bias of 0.16 to 0.35 times its SD, with coverage between 0.932 and 0.956. Orth reduces the absolute bias to at most 0.11 times its SD, with an SD at most $2\%$ larger than that of Plug-in CF and coverage between 0.944 and 0.954. The results under covariate-dependent censoring, where $G$ is estimated by survival forests, are similar to those under independent censoring. With $n=500$ (Table~\ref{tab:sim_n500}), Orth again has the smallest absolute bias among the non-oracle estimators, with near-nominal coverage.

Exponential and uniform censoring on the original time scale serve as stress tests, since with normal $\epsilon$ they violate $\E[\varphi_0^2]<\infty$ in Proposition~\ref{prop:rank-orthogonal} and Assumption~\ref{ass:cens_id}(b), respectively (Section~\ref{app:sim_sensitivity}). In these scenarios (Table~\ref{tab:sim_cens_shape}), the absolute bias of the cross-fitted estimators increases and their coverage decreases with the censoring rate, falling well below the nominal level at CR $60\%$.
Orth retains a smaller absolute bias than Plug-in CF in all six scenarios, and its mean SE stays close to the empirical SD. With heteroscedastic $\nu$ (Table~\ref{tab:sim_hetero_nu}), the results are close to those with homoscedastic $\nu$. With variance-one Laplace $\epsilon$, which has heavier tails than the normal distribution, Orth has negligible bias and near-nominal coverage under both censoring mechanisms at CR $40\%$ (Table~\ref{tab:sim_laplace}).

\begin{table}[htbp]
\centering
\caption{Simulation results under independent and covariate-dependent censoring ($n=1000$).}\label{tab:sim_results}
\small
\renewcommand{\arraystretch}{0.75}
\begin{tabular}{@{}ll rrrr rrrr@{}}
\toprule
& & \multicolumn{4}{c}{Independent} & \multicolumn{4}{c}{Covariate-dependent} \\
\cmidrule(lr){3-6} \cmidrule(lr){7-10}
CR & Method & Bias & SD & SE & CP & Bias & SD & SE & CP \\
\midrule
20\%  & Oracle        & .001 & .034 & .035 & .952 & .001 & .034 & .034 & .950 \\
      & Linear AFT    & .050 & .046 & .043 & .776 & .049 & .044 & .042 & .782 \\
      & Plug-in       & $-$.081 & .045 & .038 & .438 & $-$.082 & .047 & .037 & .412 \\
      & Plug-in CF    & $-$.014 & .040 & .041 & .936 & $-$.013 & .040 & .040 & .932 \\
      & \textbf{Orth} & $-$.005 & .041 & .041 & .944 & $-$.004 & .040 & .041 & .948 \\
\midrule
40\%  & Oracle        & .001 & .039 & .039 & .960 & .001 & .037 & .038 & .948 \\
      & Linear AFT    & .051 & .050 & .047 & .790 & .051 & .048 & .045 & .814 \\
      & Plug-in       & $-$.075 & .049 & .043 & .552 & $-$.077 & .047 & .041 & .518 \\
      & Plug-in CF    & $-$.014 & .045 & .046 & .942 & $-$.011 & .043 & .044 & .944 \\
      & \textbf{Orth} & $-$.003 & .046 & .047 & .952 & .000 & .044 & .045 & .950 \\
\midrule
60\%  & Oracle        & .001 & .047 & .046 & .940 & .002 & .044 & .044 & .954 \\
      & Linear AFT    & .051 & .059 & .055 & .844 & .053 & .055 & .052 & .816 \\
      & Plug-in       & $-$.064 & .057 & .051 & .714 & $-$.064 & .054 & .048 & .696 \\
      & Plug-in CF    & $-$.011 & .054 & .054 & .952 & $-$.008 & .051 & .052 & .956 \\
      & \textbf{Orth} & .001 & .055 & .055 & .952 & .005 & .052 & .053 & .954 \\
\bottomrule
\end{tabular}
\smallskip
\begin{flushleft}
{\footnotesize Bias, $\bar{\hat\beta}-\beta_0$; SD, empirical standard deviation; SE, mean estimated standard error; CP, 95\% CI coverage.
Oracle, nonorthogonal moment with the true $\ell_0$ and $m_0$; Linear AFT, log-normal AFT model with $X$ and $\Zbf$ entered linearly, with sandwich SE; Plug-in, nonorthogonal moment with ML nuisances and no cross-fitting; Plug-in CF, the same moment with block-pairwise cross-fitting; Orth, orthogonalized moment with block-pairwise cross-fitting. 500 replications.}
\end{flushleft}
\end{table}

\section{Real Data Application}\label{sec:app}

We apply the proposed method to electronic health record data from adults with diabetes in the All of Us Research Program \citep{the_all_of_us_research_program_investigators_all_2019}. The analysis focuses on patients with chronic kidney disease (CKD) stage 3b or worse before the first recorded evidence of diabetes (the index date). To ensure at least 4 years of potential follow-up, we restrict the cohort to index dates on or before December 31, 2020. The outcome is the time from the index date to a composite cardiovascular outcome, the first of myocardial infarction, ischemic stroke and heart failure. Patients without the outcome, including those who die, are censored at their last clinical record. The cohort comprises 1,209 patients with 545 events and a median [IQR] follow-up of 5.9 [3.4, 9.2] years.

We consider three exposures, serum albumin, estimated glomerular filtration rate (eGFR) and systolic blood pressure (SBP), which are routinely measured markers of cardiovascular risk in CKD. Each exposure ($X$) is the most recent value on or before the index date, standardized within the analysis sample of patients with that measurement. These samples comprise 1,015 patients for serum albumin, 1,062 for eGFR and 862 for SBP, with censoring of $56.7\%$, $56.3\%$ and $60.7\%$. We adjust for $q=55$, $53$ and $57$ covariates ($\Zbf$), including age at the index date, gender, race, education, laboratory measurements, comorbidity burden, and medication use. Section~\ref{app:data-application} provides details of cohort construction, outcome and exposure definitions, covariates, missing-data handling, and implementation. Table~\ref{tab:app-cohort-characteristics} summarizes cohort characteristics.

We compare seven estimators with different distributional and covariate-adjustment assumptions. The three parametric AFT benchmarks impose a normal, logistic or extreme-value error distribution and linear adjustment for $\Zbf$. Two semiparametric AFT benchmarks, a least-squares (LS) estimator and a rank-based estimator using Gehan weights \citep{chiou_fitting_2014}, leave the form of the error distribution unspecified but retain linear adjustment. The plug-in estimator solves~\eqref{eqn:rank-U-statistics} with ML nuisance estimates but without cross-fitting, allowing flexible adjustment for $\Zbf$. The proposed (Orth) estimator further applies the orthogonal correction with block-pairwise cross-fitting to remove the first-order bias from ML nuisance estimation. These estimators rest on different working models and estimating equations and need not target the same parameter under misspecification, so we interpret differences among them as sensitivity to modeling and estimation choices. Nuisance estimation follows the covariate-dependent censoring scenario of Section~\ref{subsec:setup}, with $G_0$ and $S^T_0$ estimated by survival forests, and Orth is aggregated over 10 random partitions by the median (Section~\ref{app:data-implementation}).

\begin{figure}[htbp]
    \centering
    \includegraphics[width=0.72\textwidth]{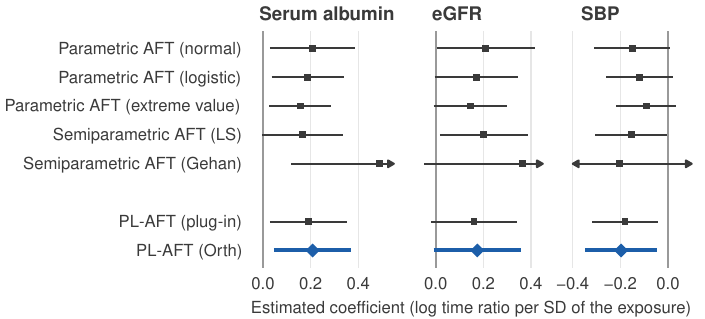}
    \caption{Estimated coefficients and 95\% confidence intervals per standard deviation of each exposure on log time to the composite cardiovascular outcome in the CKD stage 3b-or-worse cohort. Parametric AFT denotes linear AFT models with normal, logistic and extreme-value errors. Semiparametric AFT denotes the least-squares (LS) and Gehan rank estimators for the linear AFT model. PL-AFT (plug-in) and PL-AFT (Orth) denote the plug-in and proposed estimators. Intervals beyond the plotted range end in arrows. Table~\ref{tab:app-estimates} gives the numerical values.}
    \label{fig:app-dkd3b}
\end{figure}

Figure~\ref{fig:app-dkd3b} shows that all estimators agree in direction within each exposure. Higher serum albumin and eGFR are associated with later cardiovascular events, and higher SBP with earlier events, consistent with previous evidence \citep{chien_critical_2017,go_chronic_2004,rahman_central_2018}. The benchmark estimates vary across error-distribution assumptions and estimating equations. The parametric AFT estimates decrease in magnitude from the normal to the extreme-value error distribution, from 0.21 to 0.16 for serum albumin, 0.21 to 0.15 for eGFR and $-0.15$ to $-0.09$ for SBP. The LS and Gehan estimates differ most for serum albumin (0.17 versus 0.49) and eGFR (0.20 versus 0.37). The Orth estimates for serum albumin, eGFR and SBP are 0.21 (95\% CI 0.05 to 0.37), 0.17 ($-0.01$ to 0.36) and $-0.20$ ($-0.35$ to $-0.05$), and exceed the plug-in estimates in magnitude by 0.01 to 0.02. Varying the threshold that determines the horizon $U$ changes these estimates by less than 0.02 standard errors on average over the 10 partitions (Section~\ref{app:data-sensitivity}). In all three analyses, the PL-AFT estimates have standard errors comparable to those of the parametric AFT and LS estimates and less than half those of the Gehan estimates. The severity-restricted cohort of patients with CKD stage 4 or worse shows the same pattern (Figure~\ref{fig:app-dkd4}). Among the seven estimators, only Orth provides asymptotically valid inference under the conditions of Section~\ref{sec:inference} while imposing neither a parametric error distribution nor linear adjustment for $\Zbf$.

\section{Discussion} \label{sec:discussion}

We propose a rank-based DML U-statistic estimator for the partially linear AFT model, incorporating a censoring-corrected influence function for right-censored outcomes. The estimator retains the direct time-scale interpretation of the AFT model while adjusting flexibly for high-dimensional covariates. Under the conditions of Section~\ref{sec:inference}, Neyman orthogonality and block-pairwise cross-fitting yield asymptotically valid inference while allowing nuisance functions to be estimated via ML. In the primary simulation study, the estimator attains near-nominal coverage under independent and covariate-dependent Gumbel censoring at rates up to $60\%$, and the approach is illustrated in an All of Us cohort analysis of three clinical exposures and incident cardiovascular outcomes among patients with diabetes and advanced kidney disease. The resulting estimates describe associations rather than causal effects. A causal interpretation would require additional identifying assumptions, which DML does not supply \citep{fuhr_estimating_2024}.

The practical performance of the estimator is governed by nuisance estimation, particularly for $\ell_0=\E[y(T)\given\Zbf]$ under censoring.
As the censoring rate increases, this becomes harder, and the rate conditions in Assumption~\ref{ass:rates-L2} may be difficult to satisfy. The stress tests with exponential and uniform censoring illustrate the importance of information on the upper tail of $y(T)$. Future work may improve nuisance estimation under heavy censoring, allow the distribution of $\epsilon$ to depend on $(X,\Zbf)$, and extend the framework to competing risks or longitudinal covariates.

\section*{Supplementary Material}
The Supplementary Material contains simulation and application details, additional figures and tables, and proofs of Lemmas~\ref{lem:induced-smoothing}--\ref{lem:first-step} and Propositions~\ref{prop:rank-orthogonal}--\ref{prop:asymp-normal}.

\section*{Acknowledgments}
We gratefully acknowledge \textit{All of Us} participants for their contributions, without whom this research would not have been possible. We also thank the National Institutes of Health's \textit{All of Us} Research Program for making available the participant data examined in this study. This work was supported by the National Institutes of Health under grants P30 CA-16042 (GL), UL1TR000124-02 (GL), R35 GM141798 (HZ), R01 HG006139 (JZ and HZ), R01 DK142026 (GL, JZ, and HZ), and U24DK097771 (JZ and HZ), and by the National Science Foundation under grants DMS 2054253 (JZ and HZ) and IIS 2205441 (SZ, GL, JZ, and HZ).

\section*{Data and Code Availability}
This study used data from the \textit{All of Us} Research Program's Controlled Tier Dataset v9, available to authorized users on the Researcher Workbench. R code implementing the proposed method and reproducing the simulation study is available at \url{https://github.com/Tomoki-Okuno/dml-plaft}.

\section*{Competing Interests}
The authors have no relevant financial or non-financial interests to disclose.

\bibliographystyle{spbasic-doi}
\bibliography{references_zotero,references}

\clearpage
\setcounter{section}{0}
\setcounter{equation}{0}
\setcounter{table}{0}
\setcounter{figure}{0}
\setcounter{lemma}{0}
\setcounter{ass}{0}
\setcounter{remark}{0}
\renewcommand{\thesection}{S\arabic{section}}
\renewcommand{\theequation}{S\arabic{section}.\arabic{equation}}
\renewcommand{\thetable}{S\arabic{table}}
\renewcommand{\thefigure}{S\arabic{figure}}
\renewcommand{\thelemma}{S\arabic{lemma}}
\renewcommand{\theass}{S\arabic{ass}}
\renewcommand{\theremark}{S\arabic{remark}}

\begin{center}
{\Large\bfseries Supplementary Material for\\[2pt] ``Debiased machine learning for partially linear accelerated failure time models''}\\[10pt]
Tomoki Okuno, Sijie Zheng, Brendon Chau, Gang Li, Jin J. Zhou and Hua Zhou
\end{center}
\bigskip

\section{Simulation Details}\label{app:sim_details}

\subsection{Covariate generation}\label{app:covariate_generation}
Covariates are generated via a Gaussian copula with a block first-order autoregressive structure. We first draw a latent Gaussian vector $\Zbf^*\sim\mathcal{N}_{50}(0,\Sigmabf)$, where $\Sigmabf$ is block-diagonal with five $10\times 10$ blocks, each having entries $\Sigma_{jk}=0.5^{|j-k|}$. For each component $j$, the probability integral transform $\Phi(Z^*_{(j)})$ is rescaled to $[-2,2]$; the transformed columns are then centered and scaled to form the covariate vector $\Zbf$. Parentheses in $Z^*_{(j)}$ denote the $j$th component. The resulting covariates are marginally uniform with moderate within-block dependence.

\subsection{Censoring mechanisms}\label{app:sim_censoring}
Within each censoring mechanism of Section~\ref{subsec:setup}, $\sigma_C$ is held fixed across censoring rates, and $\mu_C$ is calibrated to the target censoring rate on a pilot sample of $2\times10^5$ observations. Replicate $r$ is generated from the same random seed in every scenario. Table~\ref{tab:sim_cens_quartile} reports the censoring rate within each quartile of $y(T)$. At CR $60\%$, the event is observed for $15\%$ and $12\%$ of the units in the upper quartile under independent and covariate-dependent censoring, respectively.

\begin{table}[!htbp]
\centering
\caption{Censoring rate within each quartile of $y(T)$.}\label{tab:sim_cens_quartile}
\small
\renewcommand{\arraystretch}{0.8}
\begin{tabular}{@{}ll rrrr@{}}
\toprule
& & \multicolumn{4}{c}{Quartile of $y(T)$} \\
\cmidrule(lr){3-6}
Mechanism & CR & 1st & 2nd & 3rd & 4th \\
\midrule
Independent & 20\% & .01 & .07 & .22 & .50 \\
            & 40\% & .08 & .29 & .50 & .72 \\
            & 60\% & .28 & .55 & .71 & .85 \\
\midrule
Covariate-dependent & 20\% & .00 & .06 & .21 & .52 \\
                    & 40\% & .06 & .27 & .51 & .76 \\
                    & 60\% & .24 & .54 & .73 & .88 \\
\bottomrule
\end{tabular}
\smallskip
\begin{flushleft}
{\footnotesize Monte Carlo proportions from an independent sample of $4\times10^5$ observations per scenario.}
\end{flushleft}
\end{table}

\subsection{Sensitivity scenarios}\label{app:sim_sensitivity}
The scenarios of Table~\ref{tab:sim_results} are also run with $n=500$. Three further sets of scenarios are used, the first two with independent censoring. In the first, $C$ follows an exponential distribution or a uniform distribution on $(0,u_C)$ on the original time scale, with the rate or $u_C$ calibrated to CR $\in\{20\%,40\%,60\%\}$. Both give $\log C$ a lighter right tail than the Gumbel mechanism and serve as stress tests outside our assumptions. Uniform censoring violates Assumption~\ref{ass:cens_id}(b). Under exponential censoring with rate $\lambda_C$, $G_0(u)=\exp\{-\lambda_C\exp(u)\}$ decays faster than the normal survival function $S^T_0$, so the variance of $\rho^\star_{U,0}$ given $(X,\Zbf)$ in Section~\ref{app:G-orthogonality} diverges as $U\to\infty$. Hence $\E[\varphi_0^2]=\infty$, since otherwise this variance would be bounded by $\V(\rho_0\given X,\Zbf)$ for every $U$.
These scenarios probe the truncation at the horizons of $\rho^\star_U$ controlled by Assumption~\ref{ass:tail-window}.
In the second, $\nu\given\Zbf\sim\mathcal N\{0,\sigma_\nu^2(\Zbf)\}$, as allowed by Assumption~\ref{ass:basic_id}, with $\sigma_\nu(\Zbf)\propto0.8+0.4|Z_{(2)}|+0.4|Z_{(3)}|$ rescaled in each sample to mean square one, under Gumbel censoring at the heaviest rate, CR $60\%$.
In the third, $\epsilon$ follows a Laplace distribution with variance one, under both Gumbel mechanisms at CR $40\%$. Its exponential tails are compatible with the truncation conditions in Assumption~\ref{ass:tail-window}(a) under these mechanisms.
All other settings follow Section~\ref{subsec:setup}.

\subsection{Nuisance estimation}\label{app:nuisance_estimation}
All nuisance functions are estimated for each of the $L=K(2K-1)=15$ cross-fitting blocks using only out-of-block observations (Section~\ref{subsec:cross-fitting}). The estimated correction is the horizon-specific $\varphi^\star_{U_{n,l}}$ of Section~\ref{app:G-orthogonality}, evaluated at $\hat\eta^{(l)}$, and the window $[L_{n,l},U_{n,l}]$ is the one entering Assumption~\ref{ass:tail-window}. Writing $\mathcal T_{n,l}$ for the grid of observed values of $\tilde y$ in the training sample and $\hat G^{\mathrm{KM}}_{-l}$ for the unfloored reverse Kaplan--Meier estimate from the same sample, the horizon is
\[
U_{n,l}=\max\{t\in\mathcal T_{n,l}:\hat G^{\mathrm{KM}}_{-l}(t)\ge\delta_n\},
\qquad
\delta_n=\delta_0\wedge n^{-\tau},
\]
and $L_{n,l}$ is placed below the smallest observation of the training sample. The same rule is used whether $G_0$ is estimated by the marginal Kaplan--Meier estimator or by a conditional learner. Under covariate-dependent censoring, $\hat G^{\mathrm{KM}}_{-l}$ serves only as this selection rule and does not estimate $G_0$. Defining $U_{n,l}$ on the grid rather than as a supremum over the real line matters because a step-function estimate may plateau at a positive value. Because $\hat G^{(l)}$ and $\hat S^{T,(l)}$ are step functions, all integrals in $\varphi^\star_{U_{n,l}}$ are evaluated exactly by piecewise summation over their jump points augmented by the evaluation time $\tilde y$, $U_{n,l}$, and any query point used in the tail integral, so no separate numerical approximation error arises. Within $\varphi^\star_{U_{n,l}}$, $\hat G^{(l)}$ is floored at $\delta_n$, and the floored estimate is used in the point evaluations, in the denominators, and in the hazard increments of~\eqref{eq:pi-hazard}. In the simulations $\delta_0=0.01$ and $\tau=1/2$, so $\delta_n=\delta_0$ for the reported sample sizes.

We estimate $m_0(\Zbf)$ using XGBoost regression from the \texttt{xgboost} R package \citep{chen_xgboost_2016} with squared-error objective (\texttt{reg:squarederror}), learning rate 0.02, maximum depth 1, minimum child weight 10, subsampling rate 0.7, column subsampling rate 0.8, and $L_2$ and $L_1$ penalties 10 and 0.25. In each fit, the number of boosting rounds is chosen by five-fold cross-validation within the training sample, with early stopping after 30 rounds without improvement. For $\ell_0(\Zbf)$, the same regression with maximum depth 2 and minimum child weight 50 is applied to the pseudo-outcome $\rho^\star_U$ of Section~\ref{app:G-orthogonality}, which satisfies $\E[\rho^\star_U(G_0,S^T_0)\given\Zbf]=\E[y(T)\wedge U\given\Zbf]=\ell_U(\Zbf)$. The pseudo-outcome uses $(X,\Zbf)$ through $(\hat G,\hat S^T)$, and $X$ is dropped only in the regression. Within block $l$, $\mathcal I_{-l}$ is split at random into $K_{\mathrm{in}}=5$ inner folds. The pseudo-outcomes of fold $k$ use $(\hat G,\hat S^T)$ fitted on $\mathcal I_{-l}$ without fold $k$, with the horizon given by the rule above on the same sample, and $\hat\ell^{(l)}$ is the regression fitted on all of $\mathcal I_{-l}$. The predictions of $(\ell_0,m_0)$ in the pseudo-outcomes~\eqref{eqn:pseudo-outcome} and in the pilot $\hat\beta^{(0,l)}$ are cross-fitted on the same folds. For fold $k$, $\hat m$ is fitted on $\mathcal I_{-l}$ without fold $k$, and $\hat\ell$ on pseudo-outcomes that do not use fold $k$, those of fold $k'$ being computed with $(\hat G,\hat S^T)$ and the horizon from $\mathcal I_{-l}$ without folds $k$ and $k'$. The plug-in estimator applies the same learners to the full sample without cross-fitting. Sufficient conditions for Assumption~\ref{ass:rates-L2}(i) under this construction are given after Assumption~\ref{ass:tail-window}. Apart from the number of rounds, XGBoost hyperparameters are fixed across simulation settings and are not tuned.

The conditional event survival function $S^T_0(\cdot\given X,\Zbf)$ is estimated by the survival forest of the \texttt{grf} R package \citep{athey_generalized_2019} on the 51 inputs $(X,\Zbf)$, with 500 trees, subsampling rate 0.8, 28 candidate variables per split, minimum node size 15, approximate log-rank splitting, Nelson--Aalen-type predictions and no honest splitting. Under independent censoring, the censoring survival function $G_0$ is estimated by the marginal Kaplan--Meier estimator. Under covariate-dependent censoring, $G_0(\cdot\given X,\Zbf)$ is estimated by the same forest configuration except that honest splitting is used and the censoring indicator is treated as the event indicator. The honesty and subsampling settings are chosen by comparing the integrated squared error against the true $S^T_0$ and $G_0$ in preliminary runs.

Because $\alpha_{\bullet,0}$ captures the sensitivity of the estimating equation to nuisance perturbations rather than the primary signal, the signal-to-noise ratio in its regression can be low, with $\V[\alpha_{\bullet,0}(\Zbf)]$ small relative to the noise in the pseudo-outcomes~\eqref{eqn:pseudo-outcome}. We therefore replace them, for $i\in\mathcal I_{-l}$, by the Rao--Blackwellized pseudo-outcomes of Section~\ref{subsec:cross-fitting},
\begin{equation}
\bar T^{\bullet,(l)}_i
=\iint \hat T^{\bullet,(l)}_i(u,s)\,\{-d\hat S^{\epsilon,(l)}(u)\}\,\{-d\hat G^{(l)}(s\given X_i,\Zbf_i)\},
\label{eq:rb-pseudo-outcome}
\end{equation}
where $\hat T^{\bullet,(l)}_i(u,s)$ is $\hat T^{\bullet,(l)}_i$ in~\eqref{eqn:pseudo-outcome} with $(\tilde y_i,\Delta_i)$ replaced by $((\hat\mu_i+u)\wedge s,\I\{\hat\mu_i+u\le s\})$ for $\hat\mu_i=\hat\ell(\Zbf_i)+\{X_i-\hat m(\Zbf_i)\}\hat\beta^{(0,l)}$, and $\hat S^{\epsilon,(l)}$ is the Kaplan--Meier estimate from $\{e_k(\hat\beta^{(0,l)},\hat\ell,\hat m),\Delta_k\}_{k\in\mathcal I_{-l}}$, both with the predictions $(\hat\ell,\hat m)$ cross-fitted within $\mathcal I_{-l}$. Under Assumptions~\ref{ass:basic_id}(a) and~\ref{ass:cens_id}(a), $\epsilon\perp(X,\Zbf,C)$. At the truth, $\bar T^{\bullet,(l)}_i$ is therefore the conditional expectation of $\hat T^{\bullet,(l)}_i$ given $(X_i,\Zbf_i)$ and the partners, and the residuals are $\epsilon_k\wedge c_k$ with $c_k$ independent of $\epsilon_k$ (Section~\ref{app:gehan-unbiased}), so the Kaplan--Meier estimate targets the survival function $S^\epsilon$ of $\epsilon$. Evaluated at the truth in each censoring scenario of Table~\ref{tab:sim_results}, Rao--Blackwellization with the Monte Carlo approximation below reduces the pseudo-outcome variance by 20--39\% for $\alpha_\ell$ and 56--69\% for $\alpha_m$. Under covariate-dependent censoring, $\hat G^{(l)}(\cdot\given X_i,\Zbf_i)$ in~\eqref{eq:rb-pseudo-outcome} is the out-of-bag prediction of the forest fitted on $\mathcal I_{-l}$, computed from the trees grown without unit $i$. Tail mass left after the last jump of $\hat S^{\epsilon,(l)}$ or $\hat G^{(l)}$ is placed at $+\infty$, with the pairwise weights evaluated by their limits on the extended real line. We approximate~\eqref{eq:rb-pseudo-outcome} by Monte Carlo, averaging $2\hat W^{(l)}_{ij,\bullet}$, with $(\tilde y_i,\Delta_i)$ replaced as above, over 2,000 independent draws of $(u,s,j)$ from $\hat S^{\epsilon,(l)}$, $\hat G^{(l)}(\cdot\given X_i,\Zbf_i)$ and the uniform distribution on $\mathcal J_{-l}(i)$. The approximation is conditionally unbiased given the fitted laws, and Assumption~\ref{ass:rates-L2}, imposed on the resulting $\hat\alpha^{(l)}_\bullet$, covers the errors of $\hat S^{\epsilon,(l)}$, $\hat G^{(l)}$ and the approximation.

The Riesz representers $\alpha_{\ell,0}$ and $\alpha_{m,0}$ are estimated by a generalized additive model in $\hat\ell(\Zbf)$ and $\hat m(\Zbf)$, with a thin-plate regression spline of basis dimension 8 for each and smoothing parameters selected by restricted maximum likelihood \citep{wood_fast_2011}, followed by the XGBoost regression used for $\ell_0$ fitted to its residuals on $\Zbf$. Its inputs $(\hat\ell,\hat m)$ are the predictions cross-fitted within $\mathcal I_{-l}$ for $i\in\mathcal I_{-l}$ and $(\hat\ell^{(l)},\hat m^{(l)})$ for $k\in\mathcal I_l$. We also residualize $\bar T^{\ell,(l)}_i$ on an intercept and $X_i-\hat m(\Zbf_i)$ by least squares before the regression. At the truth, this leaves the target unchanged, because $\E[\nu\given\Zbf]=0$ and, by the antisymmetry of $W_{ij,\ell}$, $\E[\alpha_{\ell,0}(\Zbf)]=0$.

\section{Additional Simulation Results}\label{app:sim_results}

Table~\ref{tab:sim_n500} reports the scenarios of Table~\ref{tab:sim_results} with $n=500$, Table~\ref{tab:sim_cens_shape} the scenarios with exponential and uniform censoring, Table~\ref{tab:sim_hetero_nu} the scenario with heteroscedastic $\nu$, and Table~\ref{tab:sim_laplace} the scenarios with Laplace $\epsilon$ (Section~\ref{app:sim_sensitivity}).

\begin{table}[!htbp]
\centering
\caption{Simulation results under independent and covariate-dependent censoring ($n=500$, 500 replications).}\label{tab:sim_n500}
\small
\renewcommand{\arraystretch}{0.8}
\begin{tabular}{@{}ll rrrr rrrr@{}}
\toprule
& & \multicolumn{4}{c}{Independent} & \multicolumn{4}{c}{Covariate-dependent} \\
\cmidrule(lr){3-6} \cmidrule(lr){7-10}
CR & Method & Bias & SD & SE & CP & Bias & SD & SE & CP \\
\midrule
20\%  & Oracle        & .000 & .050 & .050 & .946 & .000 & .050 & .049 & .934 \\
      & Linear AFT    & .052 & .068 & .060 & .822 & .051 & .066 & .059 & .836 \\
      & Plug-in       & $-$.081 & .065 & .055 & .666 & $-$.080 & .064 & .054 & .662 \\
      & Plug-in CF    & $-$.015 & .061 & .059 & .930 & $-$.014 & .060 & .058 & .934 \\
      & \textbf{Orth} & $-$.001 & .062 & .060 & .940 & $-$.001 & .061 & .059 & .930 \\
\midrule
40\%  & Oracle        & $-$.002 & .057 & .056 & .932 & .000 & .056 & .054 & .940 \\
      & Linear AFT    & .050 & .076 & .067 & .844 & .052 & .074 & .064 & .828 \\
      & Plug-in       & $-$.083 & .070 & .062 & .668 & $-$.082 & .069 & .059 & .676 \\
      & Plug-in CF    & $-$.016 & .069 & .066 & .940 & $-$.011 & .067 & .064 & .944 \\
      & \textbf{Orth} & .000 & .070 & .067 & .954 & .004 & .068 & .065 & .944 \\
\midrule
60\%  & Oracle        & $-$.004 & .067 & .065 & .940 & $-$.003 & .065 & .062 & .936 \\
      & Linear AFT    & .050 & .089 & .077 & .852 & .051 & .085 & .073 & .836 \\
      & Plug-in       & $-$.079 & .084 & .072 & .754 & $-$.078 & .083 & .069 & .718 \\
      & Plug-in CF    & $-$.012 & .081 & .078 & .930 & $-$.008 & .079 & .074 & .938 \\
      & \textbf{Orth} & .004 & .082 & .080 & .944 & .007 & .080 & .076 & .942 \\
\bottomrule
\end{tabular}
\smallskip
\begin{flushleft}
{\footnotesize Bias, $\bar{\hat\beta}-\beta_0$; SD, empirical standard deviation; SE, mean estimated standard error; CP, 95\% CI coverage. Oracle, nonorthogonal moment with the true $\ell_0$ and $m_0$; Linear AFT, log-normal AFT model with $X$ and $\Zbf$ entered linearly, with sandwich SE; Plug-in, nonorthogonal moment with ML nuisances and no cross-fitting; Plug-in CF, the same moment with block-pairwise cross-fitting; Orth, orthogonalized moment with block-pairwise cross-fitting.}
\end{flushleft}
\end{table}

\begin{table}[!htbp]
\centering
\caption{Simulation results under independent exponential and uniform censoring on the original time scale ($n=1000$, 500 replications).}\label{tab:sim_cens_shape}
\small
\renewcommand{\arraystretch}{0.8}
\begin{tabular}{@{}ll rrrr rrrr@{}}
\toprule
& & \multicolumn{4}{c}{Exponential censoring} & \multicolumn{4}{c}{Uniform censoring} \\
\cmidrule(lr){3-6} \cmidrule(lr){7-10}
CR & Method & Bias & SD & SE & CP & Bias & SD & SE & CP \\
\midrule
20\% & Oracle        & .001 & .035 & .035 & .956 & .001 & .035 & .035 & .948 \\
     & Linear AFT    & .050 & .045 & .043 & .766 & .050 & .046 & .043 & .790 \\
     & Plug-in       & $-$.085 & .047 & .038 & .402 & $-$.087 & .045 & .038 & .384 \\
     & Plug-in CF    & $-$.014 & .041 & .041 & .942 & $-$.015 & .041 & .041 & .930 \\
     & \textbf{Orth} & $-$.005 & .041 & .042 & .950 & $-$.006 & .041 & .041 & .952 \\
\midrule
40\% & Oracle        & .001 & .039 & .039 & .944 & .002 & .039 & .039 & .948 \\
     & Linear AFT    & .049 & .051 & .048 & .800 & .049 & .052 & .048 & .826 \\
     & Plug-in       & $-$.089 & .051 & .042 & .482 & $-$.099 & .048 & .042 & .394 \\
     & Plug-in CF    & $-$.022 & .045 & .046 & .926 & $-$.029 & .046 & .046 & .890 \\
     & \textbf{Orth} & $-$.011 & .045 & .047 & .942 & $-$.021 & .046 & .046 & .924 \\
\midrule
60\% & Oracle        & .004 & .046 & .046 & .934 & .002 & .047 & .046 & .942 \\
     & Linear AFT    & .050 & .061 & .056 & .846 & .046 & .060 & .056 & .866 \\
     & Plug-in       & $-$.100 & .056 & .050 & .490 & $-$.122 & .056 & .051 & .352 \\
     & Plug-in CF    & $-$.040 & .054 & .054 & .864 & $-$.057 & .053 & .055 & .808 \\
     & \textbf{Orth} & $-$.027 & .054 & .054 & .906 & $-$.051 & .053 & .056 & .842 \\
\bottomrule
\end{tabular}
\smallskip
\begin{flushleft}
{\footnotesize Notation as in Table~\ref{tab:sim_n500}.}
\end{flushleft}
\end{table}

\begin{table}[!htbp]
\centering
\caption{Simulation results with homoscedastic and heteroscedastic $\nu$ under independent Gumbel censoring (CR $60\%$, $n=1000$, 500 replications).}\label{tab:sim_hetero_nu}
\small
\renewcommand{\arraystretch}{0.8}
\begin{tabular}{@{}l rrrr rrrr@{}}
\toprule
& \multicolumn{4}{c}{Homoscedastic $\nu$} & \multicolumn{4}{c}{Heteroscedastic $\nu$} \\
\cmidrule(lr){2-5} \cmidrule(lr){6-9}
Method & Bias & SD & SE & CP & Bias & SD & SE & CP \\
\midrule
Oracle        & .001 & .047 & .046 & .940 & .000 & .047 & .046 & .942 \\
Linear AFT    & .051 & .059 & .055 & .844 & .044 & .058 & .055 & .870 \\
Plug-in       & $-$.064 & .057 & .051 & .714 & $-$.069 & .056 & .051 & .690 \\
Plug-in CF    & $-$.011 & .054 & .054 & .952 & $-$.012 & .055 & .055 & .952 \\
\textbf{Orth} & .001 & .055 & .055 & .952 & .001 & .056 & .055 & .954 \\
\bottomrule
\end{tabular}
\smallskip
\begin{flushleft}
{\footnotesize Notation as in Table~\ref{tab:sim_n500}. The homoscedastic columns repeat the CR $60\%$ independent-censoring results of Table~\ref{tab:sim_results}.}
\end{flushleft}
\end{table}

\begin{table}[!htbp]
\centering
\caption{Simulation results with variance-one Laplace error $\epsilon$ under independent and covariate-dependent Gumbel censoring (CR $40\%$, $n=1000$, 500 replications).}\label{tab:sim_laplace}
\small
\renewcommand{\arraystretch}{0.8}
\begin{tabular}{@{}l rrrr rrrr@{}}
\toprule
& \multicolumn{4}{c}{Independent} & \multicolumn{4}{c}{Covariate-dependent} \\
\cmidrule(lr){2-5} \cmidrule(lr){6-9}
Method & Bias & SD & SE & CP & Bias & SD & SE & CP \\
\midrule
Oracle        & .000 & .033 & .033 & .956 & .000 & .032 & .032 & .962 \\
Linear AFT    & .051 & .048 & .047 & .800 & .053 & .045 & .045 & .808 \\
Plug-in       & $-$.080 & .048 & .040 & .480 & $-$.082 & .045 & .038 & .426 \\
Plug-in CF    & $-$.014 & .042 & .044 & .948 & $-$.012 & .040 & .042 & .954 \\
\textbf{Orth} & $-$.003 & .043 & .044 & .946 & $-$.001 & .040 & .043 & .950 \\
\bottomrule
\end{tabular}
\smallskip
\begin{flushleft}
{\footnotesize Notation as in Table~\ref{tab:sim_n500}.}
\end{flushleft}
\end{table}

\section{Real Data Application Details}\label{app:data-application}

The \textit{All of Us} Institutional Review Board determined that use of Controlled Tier data on the Researcher Workbench does not constitute human subjects research and does not require project-specific IRB review (protocol no.~2021-02-TN-001, 17 March 2021).

\subsection{Cohort definition}\label{app:data-cohort}

The index date, the first recorded evidence of diabetes, is the earliest of the first diabetes diagnosis code (ICD-10-CM E10, E11 or E13, ICD-9-CM 250), insulin record and non-metformin oral glucose-lowering drug record. Kidney disease phenotypes combine clinical diagnosis codes with biomarker-based evidence of persistent low estimated glomerular filtration rate (eGFR), computed by the race-free CKD-EPI 2021 equation, with stage thresholds following \citet{stevens_kdigo_2024}. Renal failure is identified by ICD-10-CM N18.5, N18.6 or N19 (ICD-9-CM 585.5, 585.6 or 586), dialysis or kidney transplantation records, or persistent eGFR $<15$~mL/min/1.73~m$^2$ over at least 45 days. Chronic kidney disease (CKD) stage 4 or worse adds N18.4 (585.4) and persistent eGFR $<30$ over at least 90 days, and CKD stage 3b or worse further adds N18.32 and persistent eGFR $<45$ over at least 90 days. Persistence requires below-threshold measurements spanning the stated period with no measurement at or above the threshold in between. The primary cohort includes patients with CKD stage 3b or worse before the index date, and the severity-restricted sensitivity cohort those with CKD stage 4 or worse, both determined from codes, records and confirming eGFR measurements dated before the index date. Both cohorts are restricted to index dates on or before December 31, 2020, at least 4 years before the data cutoff (January 1, 2025) of the All of Us Curated Data Repository version 9, and to patients aged 21 years or older at the index date with gender recorded as female, male or unknown.

\subsection{Outcome and follow-up}\label{app:data-outcome}

The outcome is the first diagnosis after the index date of myocardial infarction (ICD-10-CM I21 or I22, ICD-9-CM 410), ischemic stroke (I63, 433.x1 or 434.x1) or heart failure (I50, I11.0, I13.0 or I13.2, 428, 402.x1, 404.x1 or 404.x3). A single diagnosis in any care setting suffices, including diagnoses not linked to a visit, which account for 27--30\% of the first diagnoses. Patients with such a diagnosis on or before the index date, or without follow-up after it, are excluded. Patients without the outcome, including those who die, are censored at their last clinical activity (a visit, diagnosis, procedure or drug record), excluding All of Us program visits. In the primary cohort, 55 of the 664 censored patients have a recorded death. Treating death as censoring targets the latent cardiovascular event time under the conditional independent-censoring assumption in Assumption~\ref{ass:cens_id}(a). Flexible estimation of $G_0$ does not address potential violations of this assumption due to competing mortality. Early events are retained, with 28 of the 1,209 patients in the primary cohort having an event within 30 days of the index date and 122 within one year.

\subsection{Exposures, adjustment variables and missing data}\label{app:data-covariates}

The exposures are the most recent values on or before the index date of serum albumin, eGFR and systolic blood pressure (SBP), each standardized within its analysis sample. SBP is taken from measurements coded LOINC 8480-6 between 60 and 260~mmHg. Exposures are not imputed, and patients without such a measurement are excluded, leaving 1,015 patients for serum albumin, 1,062 for eGFR and 862 for SBP. The adjustment set $\Zbf$ contains age at the index date, gender, race, education, the most recent such values of serum albumin, eGFR and the laboratory measurements in Table~\ref{tab:app-cohort-characteristics} except the exposure, the Charlson comorbidity index (CCI) \citep{charlson_new_1987} with and without its renal component, and use of insulin, metformin, SGLT2 inhibitors and statins on or before the index date. Missing values of continuous biomarkers and CCI variables are imputed by their medians among the 84,170 patients with diabetes who meet the age, gender and follow-up criteria, with missingness indicators included as separate covariates. When eGFR is the exposure, serum creatinine and its indicator are removed from $\Zbf$, because eGFR is computed from it. After dummy coding and removal of constant or duplicate columns, $\Zbf$ has $q=55$ columns for serum albumin (54 in the CKD stage 4-or-worse cohort), 53 for eGFR and 57 for SBP.

\subsection{Implementation details}\label{app:data-implementation}

The fully parametric AFT benchmarks use normal, logistic, and extreme-value error distributions on the log-time scale, corresponding to log-normal, log-logistic, and Weibull AFT models, respectively.

The two semiparametric AFT benchmarks, a least-squares estimator and a rank-based estimator with Gehan weights, are fitted with the \texttt{aftgee} R package \citep{chiou_fitting_2014}, with standard errors from 200 resamples. They are included to isolate the effect of leaving the form of the error distribution unspecified while retaining linear adjustment for $\Zbf$, providing intermediate benchmarks between the fully parametric AFT models and the partially linear estimators. Both estimators rely on resampling-based inference, which is too costly to repeat over the simulation replications, so they are not used in the simulation study (Section~\ref{subsec:setup}). A fit on each analysis sample, however, is feasible.

For the PL-AFT estimators, we fit the plug-in estimator without cross-fitting and the proposed Orth estimator with block-pairwise cross-fitting. Because censoring may depend on covariates in the data application, the nuisance functions are estimated as in Section~\ref{app:nuisance_estimation} under covariate-dependent censoring, with the hyperparameters and $\delta_0$ of the simulations. Orth uses $K=3$ folds ($L=15$ blocks), and its Monte Carlo approximation of~\eqref{eq:rb-pseudo-outcome} uses 20,000 draws. The Orth estimate depends on the random partition, so we compute it on 10 random partitions $s=1,\dots,10$ and report $\hat\beta=\mathrm{med}_s\,\hat\beta_s$ with standard error $[\mathrm{med}_s\{\widehat{\mathrm{SE}}_s^2+(\hat\beta_s-\hat\beta)^2\}]^{1/2}$ \citep{chernozhukov_doubledebiased_2018}. In the primary cohort, the 10 partition-specific Orth estimates range from 0.184 to 0.215 for serum albumin, 0.157 to 0.187 for eGFR and $-0.208$ to $-0.189$ for SBP. For the plug-in estimator, the estimate and standard error are the medians over 10 random seeds of the nuisance learners.

\subsection{Cohort characteristics}\label{app:data-characteristics}
Table~\ref{tab:app-cohort-characteristics} summarizes the primary cohort by event status, including follow-up, demographics, the three exposures, laboratory measurements, comorbidity burden, and medication use.

\begingroup
\small
\setlength{\LTcapwidth}{\textwidth}
\renewcommand{\arraystretch}{0.8}
\setlength{\tabcolsep}{3pt}
\setlength{\LTpost}{3pt}

\begin{longtable}{@{}p{0.38\textwidth}p{0.28\textwidth}p{0.28\textwidth}@{}}
\caption{Characteristics of the All of Us primary cohort by event status.}
\label{tab:app-cohort-characteristics}\\
\toprule
 & \multicolumn{2}{c}{CKD stage 3b-or-worse} \\
\cmidrule(lr){2-3}
 & Event ($n=545$) & Censored ($n=664$) \\
\midrule
\endfirsthead

\toprule
\multicolumn{3}{@{}p{\textwidth}@{}}{\tablename~\thetable. Characteristics of the All of Us primary cohort by event status, continued.} \\
\midrule
 & \multicolumn{2}{c}{CKD stage 3b-or-worse} \\
\cmidrule(lr){2-3}
 & Event ($n=545$) & Censored ($n=664$) \\
\midrule
\endhead

\midrule
\multicolumn{3}{r}{\emph{Continued on next page}}\\
\endfoot

\bottomrule
\endlastfoot

Follow-up, years & 3.7 (1.2, 7.0) & 7.2 (5.1, 10.5) \\

Age at index date, years & 57.0 (46.8, 65.0) & 55.6 (44.9, 66.0) \\

Gender &  &  \\
\quad Female & 255 (46.8\%) & 332 (50.0\%) \\
\quad Male or unknown & 290 (53.2\%) & 332 (50.0\%) \\

Race &  &  \\
\quad Black or African American & 156 (28.6\%) & 150 (22.6\%) \\
\quad White & 206 (37.8\%) & 287 (43.2\%) \\
\quad Other or unknown & 183 (33.6\%) & 227 (34.2\%) \\

Education &  &  \\
\quad Advanced degree & 66 (12.1\%) & 106 (16.0\%) \\
\quad College graduate & 90 (16.5\%) & 136 (20.5\%) \\
\quad College one to three years & 176 (32.3\%) & 201 (30.3\%) \\
\quad High school graduate or equivalent & 118 (21.7\%) & 131 (19.7\%) \\
\quad Nine through eleven & 49 (9.0\%) & 33 (5.0\%) \\
\quad Eighth grade or less, or unknown & 46 (8.4\%) & 57 (8.6\%) \\

Serum albumin, g/dL (exposure) & 3.8 (3.3, 4.2) & 4.0 (3.6, 4.3) \\
\quad Measured (analysis sample) & 439 (80.6\%) & 576 (86.7\%) \\

eGFR, mL/min/1.73~m$^2$ (exposure) & 29.8 (9.9, 47.7) & 38.1 (12.8, 57.7) \\
\quad Measured (analysis sample) & 464 (85.1\%) & 598 (90.1\%) \\

SBP, mmHg (exposure) & 134 (120, 147) & 132 (120, 144) \\
\quad Measured (analysis sample) & 339 (62.2\%) & 523 (78.8\%) \\

Glucose, mg/dL & 115 (96, 153) & 111 (92, 142) \\
\quad Missing & 59 (10.8\%) & 52 (7.8\%) \\

HbA1c, \% & 6.0 (5.4, 6.7) & 6.0 (5.4, 6.5) \\
\quad Missing & 258 (47.3\%) & 264 (39.8\%) \\

HDL cholesterol, mg/dL & 44 (36, 58) & 44 (35, 54) \\
\quad Missing & 237 (43.5\%) & 233 (35.1\%) \\

LDL cholesterol, mg/dL & 91 (69, 116) & 99 (70, 119) \\
\quad Missing & 235 (43.1\%) & 233 (35.1\%) \\

Triglycerides, mg/dL & 134 (97, 206) & 136 (94, 203) \\
\quad Missing & 233 (42.8\%) & 228 (34.3\%) \\

Total cholesterol, mg/dL & 170 (144, 203) & 176 (145, 203) \\
\quad Missing & 227 (41.7\%) & 216 (32.5\%) \\

White blood cell count, $10^3/\mu$L & 7.3 (5.7, 9.7) & 7.0 (5.6, 9.3) \\
\quad Missing & 97 (17.8\%) & 99 (14.9\%) \\

Platelet count, $10^3/\mu$L & 210 (166, 266) & 215 (176, 264) \\
\quad Missing & 90 (16.5\%) & 87 (13.1\%) \\

Hemoglobin, g/dL & 11.5 (9.9, 13.1) & 12.2 (10.4, 13.8) \\
\quad Missing & 113 (20.7\%) & 127 (19.1\%) \\

Bicarbonate, mmol/L & 25 (22, 27) & 25 (22, 28) \\
\quad Missing & 78 (14.3\%) & 72 (10.8\%) \\

Serum creatinine, mg/dL & 2.30 (1.40, 5.98) & 1.78 (1.21, 4.46) \\
\quad Missing & 80 (14.7\%) & 64 (9.6\%) \\

Calcium, mg/dL & 9.1 (8.5, 9.6) & 9.2 (8.7, 9.7) \\
\quad Missing & 75 (13.8\%) & 68 (10.2\%) \\

Blood urea nitrogen, mg/dL & 33 (22, 52) & 27 (19, 44) \\
\quad Missing & 60 (11.0\%) & 51 (7.7\%) \\

Potassium, mmol/L & 4.4 (4.0, 5.0) & 4.3 (3.9, 4.7) \\
\quad Missing & 72 (13.2\%) & 58 (8.7\%) \\

Sodium, mmol/L & 139 (136, 141) & 139 (137, 141) \\
\quad Missing & 68 (12.5\%) & 55 (8.3\%) \\

CCI & 4 (2, 5) & 4 (3, 6) \\
\quad Missing & 0 (0.0\%) & $\le$20 \\

CCI, excluding renal disease & 2 (1, 3) & 2 (1, 4) \\
\quad Missing & 0 (0.0\%) & $\le$20 \\

Insulin use & 165 (30.3\%) & 169 (25.5\%) \\
Metformin use & 22 (4.0\%) & 47 (7.1\%) \\
SGLT2 inhibitor use & $\le$20 & $\le$20 \\
Statin use & 175 (32.1\%) & 267 (40.2\%)
\end{longtable}
\noindent{\footnotesize Continuous variables are summarized as median (IQR) of the observed values, and categorical variables are summarized as $n$ (\%). For each exposure, ``Measured'' gives the patients with a measurement on or before the index date, who form its analysis sample. For the other biomarkers and CCI variables, missingness is shown as $n$ (\%) before median imputation. Counts between 1 and 20 are combined with other categories or shown as $\le$20, as required by the All of Us Data and Statistics Dissemination Policy. CCI, Charlson comorbidity index; CKD, chronic kidney disease; eGFR, estimated glomerular filtration rate; HbA1c, glycated hemoglobin; HDL, high-density lipoprotein; IQR, interquartile range; LDL, low-density lipoprotein; SBP, systolic blood pressure; SGLT2, sodium-glucose cotransporter-2.\par}
\endgroup

\subsection{Estimates and truncation diagnostics}\label{app:data-sensitivity}
\begin{figure}[htbp]
    \centering
    \includegraphics[width=0.8\textwidth]{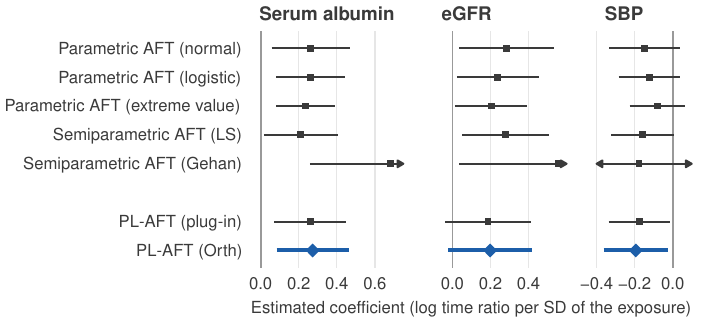}
    \caption{Estimated coefficients and 95\% confidence intervals per standard deviation of each exposure on log time to the composite cardiovascular outcome in the CKD stage 4-or-worse cohort. Methods as in Figure~\ref{fig:app-dkd3b}. Intervals beyond the plotted range end in arrows. Table~\ref{tab:app-estimates} gives the numerical values.}
    \label{fig:app-dkd4}
\end{figure}

Table~\ref{tab:app-estimates} gives the estimates in both cohorts, and Figure~\ref{fig:app-dkd4} shows those in the CKD stage 4-or-worse cohort. Table~\ref{tab:app-diagnostics} gives the analysis samples and the estimated mass of $y(T)$ beyond the horizon $U_{n,l}$ of Section~\ref{app:nuisance_estimation}. For each block, this tail mass is estimated by applying the augmented IPCW construction of $\rho^\star_U$ to $\I\{y(T)>U_{n,l}\}$ with the nuisance estimates of Orth, which is consistent when either $\hat G$ or $\hat S^T$ is, and the table reports the mean over the $L=15$ blocks. The horizons lie between 17 and 27 years after the index date, and the estimated tail mass ranges from 0.065 to 0.236.

To assess sensitivity to the horizon and the floor of $\hat G$, Orth is refitted with $\delta_0=0.005$ and $0.03$ on the same 10 partitions as the primary analysis ($\delta_0=0.01$), with $\delta_n=\delta_0$ at these sample sizes. Across the three values, the estimated tail mass ranges from 0.111 to 0.175 for serum albumin, 0.065 to 0.144 for eGFR and 0.233 to 0.284 for SBP. The Orth estimates change by less than 0.02 standard errors on average over the partitions, and the confidence intervals agree with the primary analysis on whether zero is included (Table~\ref{tab:app-estimates}).

\begin{table}[htbp]
\centering
\caption{Estimated coefficients and 95\% confidence intervals per standard deviation of each exposure on log time to the composite cardiovascular outcome.}\label{tab:app-estimates}
\footnotesize
\setlength{\tabcolsep}{3pt}
\begin{tabular}{@{}lccc@{}}
\toprule
 & Serum albumin & eGFR & SBP \\
\midrule
\multicolumn{4}{@{}l}{\emph{CKD stage 3b or worse (primary cohort)}} \\
\quad Parametric AFT (normal) & 0.207 (0.028, 0.385) & 0.209 (0.004, 0.415) & $-$0.151 ($-$0.310, 0.009) \\
\quad Parametric AFT (logistic) & 0.188 (0.036, 0.341) & 0.171 ($-$0.005, 0.346) & $-$0.120 ($-$0.260, 0.020) \\
\quad Parametric AFT (extreme value) & 0.156 (0.027, 0.286) & 0.145 ($-$0.009, 0.299) & $-$0.091 ($-$0.217, 0.036) \\
\quad Semiparametric AFT (LS) & 0.166 ($-$0.005, 0.337) & 0.202 (0.017, 0.387) & $-$0.154 ($-$0.307, $-$0.002) \\
\quad Semiparametric AFT (Gehan) & 0.488 (0.118, 0.858) & 0.366 ($-$0.049, 0.780) & $-$0.204 ($-$0.560, 0.152) \\
\quad PL-AFT (plug-in) & 0.192 (0.029, 0.354) & 0.161 ($-$0.018, 0.340) & $-$0.181 ($-$0.319, $-$0.042) \\
\quad PL-AFT (Orth) & 0.208 (0.045, 0.372) & 0.175 ($-$0.009, 0.359) & $-$0.197 ($-$0.347, $-$0.046) \\
\quad PL-AFT (Orth), $\delta_0=0.005$ & 0.209 (0.046, 0.373) & 0.175 ($-$0.007, 0.358) & $-$0.197 ($-$0.347, $-$0.046) \\
\quad PL-AFT (Orth), $\delta_0=0.03$ & 0.208 (0.044, 0.372) & 0.175 ($-$0.008, 0.358) & $-$0.197 ($-$0.347, $-$0.047) \\
\addlinespace
\multicolumn{4}{@{}l}{\emph{CKD stage 4 or worse}} \\
\quad Parametric AFT (normal) & 0.264 (0.057, 0.470) & 0.286 (0.036, 0.535) & $-$0.147 ($-$0.333, 0.038) \\
\quad Parametric AFT (logistic) & 0.262 (0.083, 0.442) & 0.238 (0.023, 0.453) & $-$0.123 ($-$0.285, 0.039) \\
\quad Parametric AFT (extreme value) & 0.238 (0.083, 0.392) & 0.204 (0.013, 0.394) & $-$0.081 ($-$0.226, 0.063) \\
\quad Semiparametric AFT (LS) & 0.211 (0.017, 0.405) & 0.279 (0.051, 0.508) & $-$0.158 ($-$0.324, 0.008) \\
\quad Semiparametric AFT (Gehan) & 0.681 (0.258, 1.103) & 0.558 (0.035, 1.080) & $-$0.178 ($-$0.632, 0.277) \\
\quad PL-AFT (plug-in) & 0.261 (0.072, 0.451) & 0.187 ($-$0.037, 0.411) & $-$0.176 ($-$0.336, $-$0.016) \\
\quad PL-AFT (Orth) & 0.273 (0.084, 0.463) & 0.198 ($-$0.023, 0.420) & $-$0.194 ($-$0.363, $-$0.024) \\
\bottomrule
\end{tabular}
\end{table}

\begin{table}[htbp]
\centering
\caption{Analysis samples, censoring and estimated tail mass beyond the horizon.}\label{tab:app-diagnostics}
\small
\begin{tabular}{@{}llrrrr@{}}
\toprule
Cohort & Exposure & $n$ & Events & Censoring (\%) & Tail mass \\
\midrule
CKD stage 3b or worse & Serum albumin & 1,015 & 439 & 56.7 & 0.126 \\
 & eGFR & 1,062 & 464 & 56.3 & 0.065 \\
 & SBP & 862 & 339 & 60.7 & 0.236 \\
\addlinespace
CKD stage 4 or worse & Serum albumin & 818 & 359 & 56.1 & 0.163 \\
 & eGFR & 836 & 370 & 55.7 & 0.096 \\
 & SBP & 680 & 273 & 59.9 & 0.216 \\
\bottomrule
\end{tabular}
\end{table}

\FloatBarrier

\section{Theoretical Derivations}\label{app:proof}

\subsection{Unbiasedness of the Gehan moment}\label{app:gehan-unbiased}

Throughout this subsection the nuisances are at the truth. Write $e_k=e_k(\beta_0,\ell_0,m_0)$, $\nu_k=X_k-m_0(\Zbf_k)$, and $\Vbf_k=(X_k,\Zbf_k)$. By \eqref{eqn:PLTM}, $y(T_k)-\ell_0(\Zbf_k)-\nu_k\beta_0=\epsilon_k$, so $e_k=\epsilon_k\wedge c_k$ and $\Delta_k=\I\{\epsilon_k\le c_k\}$, where $c_k=y(C_k)-\ell_0(\Zbf_k)-\nu_k\beta_0$. Let $S^\epsilon(t)=\prob(\epsilon>t)$ be the survival function of $\epsilon$.

\begin{lemma}\label{lem:gehan-unbiased}
Under Assumptions~\ref{ass:basic_id} and~\ref{ass:cens_id}(a), $\E[g^*(\mathcal D_i,\mathcal D_j;\beta_0,\ell_0,m_0)]=0$.
\end{lemma}

\begin{proof}
Fix $i\neq j$, write $g^*_{ij}=g^*(\mathcal D_i,\mathcal D_j;\beta_0,\ell_0,m_0)=\Delta_i\I\{e_i\le e_j\}X_{ij,0}$ with $X_{ij,0}=X_{ij}(m_0)=\nu_i-\nu_j$, and let $\mathcal H_{ij}$ be the $\sigma$-field generated by $(\Vbf_i,\Vbf_j,c_i,c_j)$. Assumption~\ref{ass:cens_id}(a) gives $\epsilon_k\perp c_k\given\Vbf_k$ and Assumption~\ref{ass:basic_id}(a) gives $\epsilon_k\perp\Vbf_k$, so $\epsilon_k\perp(\Vbf_k,c_k)$. Hence, by independence across subjects, conditionally on $\mathcal H_{ij}$, $\epsilon_i$ and $\epsilon_j$ are independent, each distributed as $\epsilon$, and
\[
\E[\Delta_i\I\{e_i\le e_j\}\given\mathcal H_{ij}]
=\prob(\epsilon_i\le c_i\wedge c_j\wedge\epsilon_j\given\mathcal H_{ij})
=\int\I\{t\le c_i\wedge c_j\}\,S^\epsilon(t-)\,\{-dS^\epsilon(t)\}.
\]
The right-hand side depends on $(c_i,c_j)$ only through $c_i\wedge c_j$ and, as $\epsilon_i$ and $\epsilon_j$ have the same law, is symmetric in $(i,j)$. Since $X_{ji,0}=-X_{ij,0}$ is $\mathcal H_{ij}$-measurable, $\E[g^*_{ij}+g^*_{ji}\given\mathcal H_{ij}]=0$. Since $\E\lvert g^*_{ij}\rvert\le\E\lvert X_{ij,0}\rvert<\infty$ by Assumption~\ref{ass:basic_id}(b) and $(\mathcal D_i,\mathcal D_j)$ is exchangeable, $\E[g^*_{ij}]=\E[g^*_{ji}]$, which gives $\E[g^*_{ij}]=0$.
\end{proof}

\subsection{Proof of Lemma~\ref{lem:induced-smoothing}}\label{app:proof-induced-smoothing}

Fix $(\beta,\ell,m)$ and write $D_{ij}=e_j(\beta,\ell,m)-e_i(\beta,\ell,m)$, $\sigma_{ij}=\Gamma_n^{1/2}\lvert X_{ij}(m)\rvert$, and $\Vbf_k=(X_k,\Zbf_k)$ for $k=i,j$. Since $\lvert\Phi(t/\sigma)-\I\{t\ge 0\}\rvert=\Phi(-\lvert t\rvert/\sigma)$ for all $t$ and $\sigma>0$,
\[
\lvert\E[\tilde g^*-g^*]\rvert
\le \E\Big[\lvert X_{ij}(m)\rvert\,\E\big\{\Delta_i\Phi\big(-\lvert D_{ij}\rvert/\sigma_{ij}\big)\given \Vbf_i,\Vbf_j\big\}\Big],
\]
where the integrand vanishes when $\sigma_{ij}=0$. Since $D_{ij}-D_{ij,0}$ is a function of $(\Vbf_i,\Vbf_j)$ and $\Delta_i$ does not depend on $(\beta,\ell,m)$, Assumption~\ref{ass:rank-M2} implies that $D_{ij}$ restricted to $\{\Delta_i=1\}$ has a conditional sub-density bounded by $L_0$, so for $\sigma_{ij}>0$,
\[
\E\big\{\Delta_i\Phi(-\lvert D_{ij}\rvert/\sigma_{ij})\given \Vbf_i,\Vbf_j\big\}
\le L_0\int\Phi(-\lvert t\rvert/\sigma_{ij})\,dt
=2L_0\sigma_{ij}\int_0^\infty\Phi(-s)\,ds
=2L_0\phi(0)\,\sigma_{ij}.
\]
Therefore $\lvert\E[\tilde g^*-g^*]\rvert\le2L_0\phi(0)\,\Gamma_n^{1/2}\,\E[X_{ij}(m)^2]\lesssim\Gamma_n^{1/2}$, where $\E[X_{ij}(m)^2]\le4\{\E[\nu^2]+\|m-m_0\|_2^2\}<\infty$ by Assumption~\ref{ass:basic_id}(b).
\qed

\subsection{Verification of \texorpdfstring{$G$}{G}-orthogonality}\label{app:G-orthogonality}

We record the condition that determines $Q^\star$ in~\eqref{eq:phi-general}, verify it for the IPCW construction, and give the finite-difference expansion used in Assumption~\ref{ass:rates-L2}(iv). Throughout, $\Vbf=(X,\Zbf)$ is fixed, $Y(u)=\I\{\tilde y\ge u\}$, and $y(T)$ is taken to have a conditional density $f_T$ given $\Vbf$ with respect to Lebesgue measure. The true censoring law $G_0$ may have atoms, and candidate survival functions $G$ and $S^T$ are right-continuous, positive where they are used, and may be discrete, so we take the censoring hazard in product-integral form,
\begin{equation}
d\Lambda^C_G(u)=-\frac{dG(u)}{G(u-)},
\qquad\text{equivalently}\qquad
d\{1/G(u)\}=\frac{1}{G(u)}\,d\Lambda^C_G(u),
\label{eq:pi-hazard}
\end{equation}
which reduces to $\Lambda^C_G=-\log G$ for continuous $G$. We perturb $G$ as $G_t=G_0+tH$ along directions $H$ of locally bounded variation satisfying $H(-\infty)=0$, $\sup_u\lvert H(u)/G_0(u)\rvert<\infty$, and $H(u)S^T_0(u)Q^\star(u)/G_0(u)\to0$ as $u\to-\infty$. We perturb $S^T$ as $S^T_t=S^T_0+tH_S$ along directions $H_S$ such that, for all small $t$, $S^T_t$ is positive with $\int_u^\infty S^T_t(s)\,ds<\infty$ for every $u$ and $\E\int\lvert Q^\star(u;\ell_0,S^T_t)\rvert\,Y(u)\,d\Lambda^C_{G_0}(u)/G_0(u)<\infty$, so that the martingale integral in~\eqref{eq:phi-general} has conditional mean zero at every $t$.

\begin{lemma}\label{lem:G-cancellation}
Suppose the $G$-sensitivity of $\xi^\star$ has the form
\[
\partial_G\E[\xi^\star(\ell_0,G_0)\given \Vbf][H]
=-\int_{-\infty}^{\infty} a^\star(u\given \Vbf)\,\frac{H(u)}{G_0(u)}\,du
\]
for some $a^\star$ with $\int_u^\infty\lvert a^\star(s\given \Vbf)\rvert\,ds<\infty$ for every $u$, that $S^T_0Q^\star$ is absolutely continuous with $S^T_0(u)Q^\star(u)\to0$ as $u\to\infty$, and that $\E\int\lvert Q^\star(u)\rvert\,Y(u)\,d\Lambda^C_{G_0}(u)/G_0(u)<\infty$, the case $t=0$ of the condition on $H_S$. Then $\partial_G\E[\varphi^\star(\ell_0,G_0,S^T_0)\given \Vbf][H]=0$ for all such $H$ if and only if
\begin{equation}
S^T_0(u\given \Vbf)\,Q^\star(u\given \Vbf)=\int_{u}^{\infty} a^\star(s\given \Vbf)\,ds .
\label{eq:Q-cancellation}
\end{equation}
\end{lemma}

\begin{proof}
Write $A(G)=\int_{-\infty}^{\infty}\{Q^\star(u)/G(u)\}\,dM^C_G(u)$. Differentiating at $G_0$ produces two terms. The term in which $\partial_G(1/G)$ acts on the integrand is the integral of the predictable $-Q^\star H/G_0^2$ against $M^C_{G_0}$, which is integrable because $\sup_u\lvert H(u)/G_0(u)\rvert<\infty$ and by the last hypothesis, and hence has conditional mean zero. The remaining term acts through the compensator. By~\eqref{eq:pi-hazard}, $\partial_G\{d\Lambda^C_{G_0}(u)\}[H]=-\{dH(u)+H(u-)\,d\Lambda^C_{G_0}(u)\}/G_0(u-)=-\{G_0(u)/G_0(u-)\}\,d\{H(u)/G_0(u)\}$, at atoms of $G_0$ as well as on its continuous part, the second equality by the product rule $d\{H/G_0\}(u)=H(u-)\,d\{1/G_0\}(u)+dH(u)/G_0(u)$. Therefore $\partial_G\{dM^C_{G_0}\}[H]=Y(u)\{G_0(u)/G_0(u-)\}\,d\{H(u)/G_0(u)\}$. Using $\E[Y(u)\given \Vbf]=\E[\I\{y(T)\ge u\}\I\{y(C)\ge u\}\given\Vbf]=S^T_0(u-)G_0(u-)=S^T_0(u)G_0(u-)$, the second equality by Assumption~\ref{ass:cens_id}(a) and the third because $y(T)$ has a density,
\[
\partial_G\E[A(G_0)\given \Vbf][H]
=\int_{-\infty}^{\infty} Q^\star(u)S^T_0(u)\,d\!\left\{\frac{H(u)}{G_0(u)}\right\}
=-\int_{-\infty}^{\infty}\frac{H(u)}{G_0(u)}\,d\{S^T_0(u)Q^\star(u)\},
\]
by integration by parts, which needs no correction for common jumps because $S^T_0Q^\star$ is absolutely continuous, the boundary terms vanishing at $-\infty$ by the direction class and at $+\infty$ by hypothesis. Adding the sensitivity of $\xi^\star$ and requiring the sum to vanish for all $H$ gives $d\{S^T_0(u)Q^\star(u)\}=-a^\star(u)\,du$, which integrates to~\eqref{eq:Q-cancellation}.
\end{proof}

For the IPCW construction,
\[
\E[\xi^{I}(\ell_0,G)\given\Vbf]
=\E\bigg[\frac{G_0(y(T)-)\{y(T)-\ell_0(\Zbf)\}}{G(y(T))}\given\Vbf\bigg]
=\int\{u-\ell_0(\Zbf)\}f_T(u)\frac{G_0(u)}{G(u)}\,du,
\]
the first equality because $\tilde y=y(T)$ on $\{\Delta=1\}$ and $\E[\Delta\given y(T),\Vbf]=G_0(y(T)-)$ by Assumption~\ref{ass:cens_id}(a), and the second because $y(T)$ has a density, so $a^{I}(u)=\{u-\ell_0(\Zbf)\}f_T(u)$. By~\eqref{eq:ipcw-residual}, $S^T_0Q^{I}=uS^T_0+\int_u^\infty S^T_0-\ell_0S^T_0$, which is absolutely continuous because $S^T_0$ is, with $d\{S^T_0Q^{I}\}=-\{u-\ell_0(\Zbf)\}f_T(u)\,du=-a^{I}(u)\,du$. Assumption~\ref{ass:cens_id}(c) gives $\int_u^\infty\lvert a^{I}\rvert<\infty$ and $S^T_0(u)Q^{I}(u)\to0$ as $u\to\infty$, so $S^T_0(u)Q^{I}(u)=\int_u^\infty a^{I}(s)\,ds$, which is~\eqref{eq:Q-cancellation} for $Q^{I}$. The right-hand side equals $\E[\{y(T)-\ell_0(\Zbf)\}\I\{y(T)\ge u\}\given\Vbf]$, and dividing by $S^T_0(u)=\prob(y(T)\ge u\given\Vbf)$ gives $Q^{I}(u)=\E[y(T)-\ell_0(\Zbf)\given y(T)\ge u,\Vbf]$.

\medskip\noindent\textbf{Horizon-specific construction.}
A step-function estimate of $S^T$ may remain positive beyond the largest observation, making $\int_u^\infty S^T(v)\,dv$, and hence the full-support $Q^{I}$ in~\eqref{eq:ipcw-residual}, infinite. We therefore evaluate the correction up to a finite horizon $U$ and take $y(T)\wedge U$ as the target at that horizon. Assume $G>0$ and $S^T>0$ on $(-\infty,U]$, write $A_{S,U}(u)=\int_u^U S^T(v)\,dv$ and $A_{S_0,U}$ for its value at $S^T_0$, and set
\begin{align}
\xi^{I}_U(\ell,G)&=\frac{\Delta\,\I\{\tilde y\le U\}\{\tilde y-\ell(\Zbf)\}}{G(\tilde y\given X,\Zbf)}
+\frac{\I\{\tilde y>U\}\{U-\ell(\Zbf)\}}{G(U\given X,\Zbf)},\notag\\
Q^{I}_{S,U}&=u+\frac{A_{S,U}(u\given x,\zbf)}{S^T(u\given x,\zbf)}-\ell(\zbf),
\label{eq:ipcw-horizon}
\end{align}
with $\varphi^\star_U(\ell,G,S^T)=\xi^{I}_U(\ell,G)+\int_{-\infty}^{U}\{Q^{I}_{S,U}(u)/G(u)\}\,dM^C_G(u)$. At the truth, $\E[\xi^{I}_U(\ell_0,G_0)\given\Zbf]=\E[y(T)\wedge U\given\Zbf]-\ell_0(\Zbf)$. At the true nuisances and under Assumption~\ref{ass:cens_id}(c), the construction converges to its full-support counterpart as $U\to\infty$.

\medskip\noindent\textbf{Additive separation of $\ell$.}
Algorithm~\ref{alg:rank} and Assumption~\ref{ass:rates-L2}(iv) use the pseudo-outcome $\rho^\star_U(G,S^T)=\varphi^\star_U(0,G,S^T)$, that is, $\varphi^\star_U$ with $\ell\equiv0$. Since $\int_{-\infty}^{U}G^{-1}dM^C_G=(1-\Delta)\I\{\tilde y\le U\}/G(\tilde y)-1/G(\tilde y\wedge U)+1$ by~\eqref{eq:pi-hazard} and $G(-\infty)=1$,
\begin{equation}
\frac{\Delta\,\I\{\tilde y\le U\}}{G(\tilde y)}+\frac{\I\{\tilde y>U\}}{G(U)}
+\int_{-\infty}^{U}\frac{dM^C_G(u)}{G(u)}=1
\qquad\text{pathwise, for every candidate }G.
\label{eq:pathwise-one}
\end{equation}
Hence the coefficient of $\ell(\Zbf)$ in $\varphi^\star_U$ is $-1$, so that, for every $\ell$,
\[
\varphi^\star_U(\ell,G,S^T)=\rho^\star_U(G,S^T)-\ell(\Zbf),
\qquad
\E[\rho^\star_U(G_0,S^T_0)\given\Zbf]=\E[y(T)\wedge U\given\Zbf].
\]

\medskip\noindent\textbf{Finite-difference expansion at a horizon.}
Write $r_G=G_0/G$ and $r_S=S^T_0/S^T$, both locally of bounded variation on $(-\infty,U]$, and $\rho^\star_{U,0}=\rho^\star_U(G_0,S^T_0)$. Taking conditional expectations term by term, with $\E[dM^C_G(u)\given\Vbf]=S^T_0(u)G_0(u-)\{d\Lambda^C_{G_0}(u)-d\Lambda^C_G(u)\}$, the product-integral identity $dr_G(u)=\{G_0(u-)/G(u)\}\{d\Lambda^C_G(u)-d\Lambda^C_{G_0}(u)\}$, and $S^T_0Q^{I}_{S,U}=uS^T_0+r_SA_{S,U}$ at $\ell\equiv0$,
\[
\begin{aligned}
&\E[\rho^\star_U(G,S^T)-\rho^\star_{U,0}\given\Vbf]\\
&\quad=\int_{-\infty}^{U}uf_T(u)\{r_G(u)-1\}\,du+US^T_0(U)\{r_G(U)-1\}\\
&\qquad-\int_{-\infty}^{U}\{uS^T_0(u)+r_S(u)A_{S,U}(u)\}\,dr_G(u)\\
&\quad=\int_{-\infty}^{U}\{r_G(u-)-1\}\,d\{r_SA_{S,U}-A_{S_0,U}\}(u)\\
&\qquad+\lim_{u\to-\infty}\{r_G(u)-1\}\{uS^T_0(u)+r_S(u)A_{S,U}(u)\}.
\end{aligned}
\]
The second equality integrates the last integral by parts, using $d\{uS^T_0+A_{S_0,U}\}=-uf_T\,du$, and the boundary term at $u=U$ cancels the horizon term $US^T_0(U)\{r_G(U)-1\}$. Since $A_{S,U}$ is absolutely continuous with $dA_{S,U}=-S^T\,du$, the product rule gives $d\{r_SA_{S,U}-A_{S_0,U}\}=A_{S,U}\,dr_S$ without a correction for common jumps. Therefore, whenever the limit vanishes,
\begin{equation}
\E[\rho^\star_U(G,S^T)-\rho^\star_{U,0}\given\Vbf]
=\int_{-\infty}^{U}\{r_G(u-)-1\}\,A_{S,U}(u)\,dr_S(u).
\label{eq:phi-double-robust}
\end{equation}
The left limit $r_G(u-)$ matters only at common jumps of $r_G$ and $r_S$, and may be replaced by $r_G(u)$ when there are none. The right-hand side vanishes when $G=G_0$ and when $S^T=S^T_0$. Averaging over $X$ given $\Zbf$ and weighting by $\alpha_{\ell,0}(\Zbf)$ gives the first of the two terms in $\E[\alpha_{\ell,0}\{\hat\rho^{(l)}-\rho_0\}]$, the second being the horizon bias $-\E[\alpha_{\ell,0}\{y(T)-U\}^+]$, where $a^+=\max\{a,0\}$.

\medskip\noindent\textbf{Window conditions for Assumption~\ref{ass:rates-L2}.}
Let $I_{n,l}=[L_{n,l},U_{n,l}]$ be the interval used on block $l$, with $U_{n,l}$ the computational horizon and $L_{n,l}$ below the smallest observation in the training sample. Write $\hat\rho^{(l)}=\rho^\star_{U_{n,l}}(\hat G^{(l)},\hat S^{T,(l)})$ for the horizon-specific estimate entering Assumption~\ref{ass:rates-L2}(iv). Splitting the integral in~\eqref{eq:phi-double-robust} over $(-\infty,L_{n,l})$ and $I_{n,l}$, and adding the horizon difference $\rho^\star_{U_{n,l},0}-\rho_0$, whose conditional mean is $-\E[\{y(T)-U_{n,l}\}^+\given \Vbf]$, gives
\begin{equation}
\E[\alpha_{\ell,0}\{\hat\rho^{(l)}-\rho_0\}]
=\underbrace{\E[\alpha_{\ell,0}B_{I_{n,l}}]}_{\text{central}}
+\underbrace{\E[\alpha_{\ell,0}B_{-,n,l}]}_{\text{lower tail}}
-\underbrace{\E[\alpha_{\ell,0}\{y(T)-U_{n,l}\}^+]}_{\text{horizon}},
\label{eq:iv-decomposition}
\end{equation}
where $B_I(\Vbf)=\int_I\{r_G(u-)-1\}A_{S,U}(u)\,dr_S(u)$ and $B_{-,n,l}$ is the same integral over $(-\infty,L_{n,l})$. The three terms are controlled separately.

\emph{Central term.} Write $\|\cdot\|_{\infty,I}$ and $\mathrm{TV}_I$ for the supremum norm and the total variation on an interval $I$, and $\|f\|_{p\text{-}\mathrm{var},I}=\sup\{\sum_{k=1}^m|f(t_k)-f(t_{k-1})|^p\}^{1/p}$ for the $p$-variation seminorm, the supremum being over finite partitions $t_0<\dots<t_m$ of $I$.

\begin{lemma}\label{lem:pvar}
Fix $p,q>1$ with $p^{-1}+q^{-1}>1$, and suppose $G>0$ and $S^T>0$ on $I=[L,U]$. Then
\begin{gather*}
|B_I(\Vbf)|\le C_{p,q}\,(U-L)\,R_G(\Vbf)\,R_S(\Vbf),\\
R_G=\|r_G(\cdot-)-1\|_{\infty,I}+\|r_G(\cdot-)-1\|_{p\text{-}\mathrm{var},I},
\quad
R_S=\|r_S-1\|_{q\text{-}\mathrm{var},I},
\end{gather*}
with $C_{p,q}$ depending only on $(p,q)$.
\end{lemma}

\begin{proof}
The Young--Stieltjes inequality \citep{dudley_differentiability_1999} gives $|\int_I f\,dh|\le c_{p,q}\{|f(L)|+\|f\|_{p\text{-}\mathrm{var},I}\}\|h\|_{q\text{-}\mathrm{var},I}$ for $f=(r_G(\cdot-)-1)A_{S,U}$ and $h=r_S-1$, with $c_{p,q}$ depending only on $(p,q)$. By the product bound for $p$-variation and $\|A_{S,U}\|_{p\text{-}\mathrm{var},I}\le\mathrm{TV}_I(A_{S,U})$, $\|f\|_{p\text{-}\mathrm{var},I}\le\|A_{S,U}\|_{\infty,I}\|r_G(\cdot-)-1\|_{p\text{-}\mathrm{var},I}+\mathrm{TV}_I(A_{S,U})\|r_G(\cdot-)-1\|_{\infty,I}$. The function $A_{S,U}$ is nonincreasing with $A_{S,U}(U)=0$, so $\|A_{S,U}\|_{\infty,I}=\mathrm{TV}_I(A_{S,U})=A_{S,U}(L)\le U-L$, uniformly over candidates $S^T$. Hence $|f(L)|+\|f\|_{p\text{-}\mathrm{var},I}\le2(U-L)R_G$, and the lemma holds with $C_{p,q}=2c_{p,q}$.
\end{proof}

Let $R_G^{(l)}$ and $R_S^{(l)}$ denote the quantities of Lemma~\ref{lem:pvar} evaluated at the block-$l$ nuisances on $I_{n,l}$, and let $\mathcal D_{-l}=\{\mathcal D_i:i\in\mathcal I_{-l}\}$. Here and below, conditioning on a training sample, such as $\mathcal D_{-l}$, also conditions on any auxiliary randomness used to construct the nuisance estimates from it. By the Cauchy--Schwarz inequality, $|\E[\alpha_{\ell,0}B_{I_{n,l}}]|\le C_{p,q}(U_{n,l}-L_{n,l})\mathcal R^{(l)}_{G,n}\mathcal R^{(l)}_{S,n}$, where
\[
\mathcal R^{(l)}_{G,n}=\Bigl[\E\bigl\{|\alpha_{\ell,0}(\Zbf)|\,R_G^{(l)}(\Vbf)^2\given\mathcal D_{-l}\bigr\}\Bigr]^{1/2},
\qquad
\mathcal R^{(l)}_{S,n}=\Bigl[\E\bigl\{|\alpha_{\ell,0}(\Zbf)|\,R_S^{(l)}(\Vbf)^2\given\mathcal D_{-l}\bigr\}\Bigr]^{1/2}.
\]
For illustration, if $\mathcal R^{(l)}_{G,n}$ and $\mathcal R^{(l)}_{S,n}$ are of orders $n^{-(p-1)/p}$ and $n^{-(q-1)/q}$, Assumption~\ref{ass:tail-window}(c) requires $U_{n,l}-L_{n,l}=o_p(n^{3/2-p^{-1}-q^{-1}})$, which is nonvacuous exactly when $p^{-1}+q^{-1}<3/2$. Together with $p^{-1}+q^{-1}>1$ in Lemma~\ref{lem:pvar}, this gives $1<p^{-1}+q^{-1}<3/2$, or $4/3<p<2$ when $p=q$. The exponents for $G$ and $S^T$ need not agree. We do not claim that the survival forests used here attain such rates, and impose the product condition at a high level.

\emph{Lower tail.} Below the smallest observation in the training sample, both $\hat G^{(l)}$ and $\hat S^{T,(l)}$ equal one, so $r_G=G_0$, $r_S=S^T_0$ and $dr_S=dS^T_0$ there. Using $A_{S,U}(u)\le U-u$,
\begin{equation}
|B_{-,n,l}(\Vbf)|
\le\int_{-\infty}^{L_{n,l}}\{1-G_0(u\given \Vbf)\}\,(U_{n,l}-u)\,\{-dS^T_0(u\given \Vbf)\},
\label{eq:lower-envelope}
\end{equation}
an envelope which, conditionally on the training sample, involves only the true conditional laws and the training-sample-measurable endpoints. Choosing $L_{n,l}$ below the smallest observation is what makes this simplification available. If $L_{n,l}$ were placed inside the data, the lower tail would carry nuisance estimation error and would require a separate stochastic bound.

\emph{Horizon.} The last term of~\eqref{eq:iv-decomposition} is the bias of the horizon-specific target. A bias of the same kind enters $\hat\ell^{(l)}$, whose pseudo-outcomes target $\ell_U(\Zbf)=\E[y(T)\wedge U\given\Zbf]$ at the horizons $U$ of their own training samples rather than $\ell_0$. Let $U_{l,k}$ be the horizon of the pseudo-outcomes of inner fold $k$ within block $l$ (Section~\ref{app:nuisance_estimation}) and $\mathcal U^\ell_n=\{U_{l,k}:l=1,\dots,L,\ k=1,\dots,K_{\mathrm{in}}\}$.

\begin{ass}[Window]\label{ass:tail-window}
The horizons satisfy $\min_lU_{n,l}\overset p\to\infty$, and there exist training-sample-measurable $L_{n,l}$ below the smallest observation of the training sample and fixed $p,q>1$ with $p^{-1}+q^{-1}>1$ such that
(a) $\max_{U\in\mathcal U^\ell_n}\|\ell_U-\ell_0\|_{2,w}=o_p(n^{-1/4})$ and $\max_l\bigl|\E[\alpha_{\ell,0}(\Zbf)\{y(T)-U_{n,l}\}^+]\bigr|=o_p(n^{-1/2})$,
(b) $\max_l\E\bigl[|\alpha_{\ell,0}|\,\lvert B_{-,n,l}\rvert\bigr]=o_p(n^{-1/2})$, for which~\eqref{eq:lower-envelope} is sufficient, and
(c) $\max_l(U_{n,l}-L_{n,l})\,\mathcal R_{G,n}^{(l)}\mathcal R_{S,n}^{(l)}=o_p(n^{-1/2})$.
All expectations are taken as in Assumption~\ref{ass:rates-L2}.
\end{ass}

\noindent
The first condition in (a) controls the truncation contribution to Assumption~\ref{ass:rates-L2}(i), and the remaining conditions control~(iv). The admissible range for $U_{n,l}-L_{n,l}$ is determined jointly by the tails of the true conditional laws and by the rates of the nuisance learners, so no single moment condition is the right primitive.

\medskip\noindent\textbf{Sufficient conditions for Assumption~\ref{ass:rates-L2}(i) in $\ell$.}
For the following sufficient conditions, assume $\rho_0\in L_2(\mathcal P)$. Let $F_{l,1},\dots,F_{l,K_{\mathrm{in}}}$ be the inner folds of $\mathcal I_{-l}$ in Section~\ref{app:nuisance_estimation}, and let $(\hat G_{l,k},\hat S^T_{l,k})$ be the fits for the pseudo-outcomes of fold $k$ and $[L_{l,k},U_{l,k}]$ their window, all computed from $\mathcal I_{-l}\setminus F_{l,k}$. Conditionally on these data, the pseudo-outcomes of fold $k$ are independent across units, with conditional mean $\ell_{U_{l,k}}(\Zbf)+\E[B_{l,k}\given\Zbf]$ given $\Zbf$, where $B_{l,k}$ is the right-hand side of~\eqref{eq:phi-double-robust} at $(\hat G_{l,k},\hat S^T_{l,k},U_{l,k})$. Let $\bar\ell^{(l)}$ be the average of these conditional means with weights $|F_{l,k}|/|\mathcal I_{-l}|$. Then
\[
\|\hat\ell^{(l)}-\ell_0\|_{2,w}
\le\|\hat\ell^{(l)}-\bar\ell^{(l)}\|_{2,w}+\max_k\|B_{l,k}\|_{2,w}+\max_k\|\ell_{U_{l,k}}-\ell_0\|_{2,w},
\]
with $\|\cdot\|_{2,w}$ as in Assumption~\ref{ass:rates-L2}(i), and the $\ell$-part of Assumption~\ref{ass:rates-L2}(i) holds if the three terms are $o_p(n^{-1/4})$ uniformly in $l$. The last term is covered by the first condition in Assumption~\ref{ass:tail-window}(a), since $U_{l,k}\in\mathcal U^\ell_n$. By Lemma~\ref{lem:pvar} and~\eqref{eq:lower-envelope}, the middle term is $o_p(n^{-1/4})$ when $\max_{l,k}(U_{l,k}-L_{l,k})\|R_GR_S\|_{2,w}=o_p(n^{-1/4})$ and $\max_{l,k}\|B_{-}\|_{2,w}=o_p(n^{-1/4})$, with $R_G$, $R_S$ and the lower-tail term $B_{-}$ evaluated at the inner fits. Unlike Assumption~\ref{ass:tail-window}(c), this norm is weighted by $w$ rather than by $\alpha_{\ell,0}$ and involves the inner fits rather than the block fits. The first term is the error of the regression learner applied to pseudo-outcomes that are independent within each of the $K_{\mathrm{in}}$ folds given the corresponding fits. This is a high-level condition on the learner. General oracle inequalities for regression with estimated outcomes are given by \citet{kennedy_towards_2023}; we do not apply them to the present construction and impose the rate directly. Such bounds typically require the conditional second moment of the pseudo-outcome to be bounded in probability, and for the oracle pseudo-outcome this holds uniformly in $U$. By~\eqref{eq:pathwise-one} at $(G_0,S^T_0)$, $\rho^\star_{U,0}=h_U-\int_{-\infty}^{U}\{h_U-\bar Q^\star_{0,U}(u)\}\,dM^C_{G_0}(u)/G_0(u)$ with $h_U=y(T)\wedge U$ and $\bar Q^\star_{0,U}(u)=\E[h_U\given y(T)\ge u,\Vbf]$. Conditionally on $(y(T),\Vbf)$, $M^C_{G_0}$ is a martingale by Assumption~\ref{ass:cens_id}(a), so the integral has conditional mean zero given $(y(T),\Vbf)$. Its predictable variation is $\int f^2\,d\langle M^C_{G_0}\rangle$ with $f=\{h_U-\bar Q^\star_{0,U}\}/G_0$, where $d\langle M^C_{G_0}\rangle(u)=Y(u)\{G_0(u)/G_0(u-)\}\,d\Lambda^C_{G_0}(u)$ and $\E[Y(u)\given y(T),\Vbf]=\I\{y(T)\ge u\}G_0(u-)$, so the factor $G_0(u-)$ cancels and
\[
\V(\rho^\star_{U,0}\given\Vbf)=\V(h_U\given\Vbf)+\int_{-\infty}^{U}\V(h_U\given y(T)\ge u,\Vbf)\,S^T_0(u)\,\frac{d\Lambda^C_{G_0}(u)}{G_0(u)}\le\V(\rho_0\given\Vbf),
\]
the inequality because truncation at $U$ is $1$-Lipschitz and the right-hand side is the same expression with $U=\infty$ and $h_\infty=y(T)$. Hence $\E[\{\rho^\star_{U,0}-\ell_U(\Zbf)\}^2]=\E\{\V(\rho^\star_{U,0}\given\Vbf)\}+\E[\{\E(h_U\given\Vbf)-\ell_U(\Zbf)\}^2]\le\E\{\V(\rho_0\given\Vbf)\}+\E\{\V(y(T)\given\Zbf)\}$, which is finite when $\rho_0\in L_2(\mathcal P)$ because $\E[y(T)^2]\le\E[\rho_0^2]$ by the same decomposition at $U=\infty$.

\subsection{Proof of Proposition~\ref{prop:rank-orthogonal}}\label{app:proof-rank-orthogonal}

\medskip\noindent\textit{G\^ateaux derivatives of $\tilde g$ and identification of $W_{ij,\bullet}$.}
We first compute the G\^ateaux derivatives of $\E[\tilde g]$ with respect to $\ell$ and $m$, establishing that the pairwise weights $W_{ij,\ell}$ and $W_{ij,m}$ defined in~\eqref{eqn:pairwise-weights} are precisely the kernels that appear.
Write $h_{\ell,ij}=h_\ell(\Zbf_i)-h_\ell(\Zbf_j)$ and $h_{m,ij}=h_m(\Zbf_i)-h_m(\Zbf_j)$ for square-integrable perturbation directions. At the truth, write $e_{k,0}=e_k(\beta_0,\ell_0,m_0)$, let $D_{ij,0}$ be as in Assumption~\ref{ass:rank-M2}, and let $E_{ij,0}$ and $X_{ij,0}$ be as in~\eqref{eqn:pairwise-weights}.

\textit{Derivative with respect to $\ell$.}
Perturbing $\ell_0\mapsto \ell_0+rh_\ell$ changes $e_k\mapsto e_{k,0}-rh_\ell(\Zbf_k)$, so $D_{ij,r}=D_{ij,0}+rh_{\ell,ij}$ while $X_{ij}(m_0)$ is unchanged. Differentiating $\E[\tilde g^*]$ at $r=0$,
\[
\partial_\ell\E[\tilde g^*(\mathcal D_i,\mathcal D_j;\beta_0,\ell_0,m_0)][h_\ell]
=\E\left[\Delta_i\,\phi(E_{ij,0})\,\frac{\mathrm{sgn}(X_{ij,0})}{\Gamma_n^{1/2}}\,h_{\ell,ij}\right].
\]
Symmetrizing over $\tilde g=\tfrac12(\tilde g^*_{ij}+\tilde g^*_{ji})$, where $\tilde g^*_{ij}=\tilde g^*(\mathcal D_i,\mathcal D_j;\beta_0,\ell,m)$, and using $E_{ji,0}=-E_{ij,0}$, $X_{ji,0}=-X_{ij,0}$, $h_{\ell,ji}=-h_{\ell,ij}$,
\begin{equation}
\partial_\ell\E[\tilde g][h_\ell]
=\E\Big[W_{ij,\ell}\,\big\{h_\ell(\Zbf_i)-h_\ell(\Zbf_j)\big\}\Big],
\label{eq:gateaux-ell}
\end{equation}
with $W_{ij,\ell}=\tfrac{\Delta_i+\Delta_j}{2}\,\phi(E_{ij,0})\,\tfrac{\mathrm{sgn}(X_{ij,0})}{\Gamma_n^{1/2}}$.

\textit{Derivative with respect to $m$.}
Perturbing $m_0\mapsto m_0+rh_m$ changes $e_k\mapsto e_{k,0}+r\beta_0h_m(\Zbf_k)$, so $D_{ij,r}=D_{ij,0}-r\beta_0h_{m,ij}$ and $X_{ij,r}=X_{ij,0}-rh_{m,ij}$. For $\tilde g^*_{ij}$, on $\{X_{ij,0}\neq0\}$ the quotient rule on $E_{ij,r}=D_{ij,r}/(\Gamma_n^{1/2}\lvert X_{ij,r}\rvert)$ yields
\[
\left.\frac{dE_{ij,r}}{dr}\right|_{r=0}
=\frac{h_{m,ij}}{\lvert X_{ij,0}\rvert}
\left(E_{ij,0}\,\mathrm{sgn}(X_{ij,0})-\frac{\beta_0}{\Gamma_n^{1/2}}\right),
\]
and the product rule applied to $\Phi(E_{ij,r})\,X_{ij,r}$ gives
\[
\partial_m\E[\tilde g^*_{ij}][h_m]
=\E\bigg[\Delta_i\,h_{m,ij}\bigg\{\phi(E_{ij,0})\Big(E_{ij,0}-\frac{\beta_0\,\mathrm{sgn}(X_{ij,0})}{\Gamma_n^{1/2}}\Big)-\Phi(E_{ij,0})\bigg\}\bigg].
\]
For $\tilde g^*_{ji}$, using $E_{ji,0}=-E_{ij,0}$, $X_{ji,0}=-X_{ij,0}$, $h_{m,ji}=-h_{m,ij}$, $\phi(-x)=\phi(x)$, and $\Phi(-x)=1-\Phi(x)$,
\[
\partial_m\E[\tilde g^*_{ji}][h_m]
=\E\bigg[\Delta_j\,h_{m,ij}\bigg\{\phi(E_{ij,0})\Big(E_{ij,0}-\frac{\beta_0\,\mathrm{sgn}(X_{ij,0})}{\Gamma_n^{1/2}}\Big)+1-\Phi(E_{ij,0})\bigg\}\bigg].
\]
Averaging gives
\begin{equation}
\partial_m\E[\tilde g][h_m]
=\E\Big[W_{ij,m}\,\big\{h_m(\Zbf_i)-h_m(\Zbf_j)\big\}\Big],
\label{eq:gateaux-m}
\end{equation}
with $W_{ij,m}=\tfrac{\Delta_i+\Delta_j}{2}\big[\phi(E_{ij,0})\big\{E_{ij,0}-\tfrac{\beta_0\,\mathrm{sgn}(X_{ij,0})}{\Gamma_n^{1/2}}\big\}-\Phi(E_{ij,0})\big]+\tfrac{\Delta_j}{2}$.

Differentiation under the expectation is justified by the representation $\E[\tilde g^*_{ij}\given\Vbf_i,\Vbf_j]=F(a,x)$ in the proof of Lemma~\ref{lem:g-remainder}, which also covers $\{X_{ij,0}=0\}$. Along the two paths, $(a_r,x_r)=(rh_{\ell,ij},X_{ij,0})$ and $(-r\beta_0h_{m,ij},X_{ij,0}-rh_{m,ij})$. By Assumption~\ref{ass:rank-M2} alone, $F$ is continuously differentiable with $|F_a|\le L_0|x|$ and $0\le F_x\le1$, so for $r\in[0,1]$ the derivative in $r$ is dominated by $L_0(|X_{ij,0}|+|h_{m,ij}|)(|h_{\ell,ij}|+|\beta_0||h_{m,ij}|)+|h_{m,ij}|$, which is integrable. On $\{X_{ij,0}=0\}$, $F_a=0$ and $F_x=\bar Q(-a)$, in agreement with $W_{ij,\ell}$ and $W_{ij,m}$ under the convention after~\eqref{eqn:rank-smooth-moment}.

\textit{Cancellation by $\gamma$.} Let $\alpha_{\bullet,0}$ be the projected sensitivity function defined following~\eqref{eqn:pairwise-weights}.
By the antisymmetry of $W_{ij,\bullet}$ and the tower property, for $\bullet\in\{\ell,m\}$,
\[
\E\big[W_{ij,\bullet}\{h(\Zbf_i)-h(\Zbf_j)\}\big]
=
\E\big[\alpha_{\bullet,0}(\Zbf)\,h(\Zbf)\big]
\]
for any square-integrable $h$, where $\Zbf$ is a generic copy.

For the $\ell$-component of $\gamma$: since $\varphi^\star=\rho^\star(G,S^T)-\ell(\Zbf)$ with $\rho^\star$ free of $\ell$ (Section~\ref{subsec:orthogonality}), $\partial_\ell\varphi^\star_k[h_\ell]=-h_\ell(\Zbf_k)$. Hence
$\partial_\ell\E[\gamma][h_\ell] = -\E[\alpha_{\ell,0}(\Zbf)\,h_\ell(\Zbf)].$
Combining with~\eqref{eq:gateaux-ell} gives $\partial_\ell\E[\tilde\psi][h_\ell]=\E[\alpha_{\ell,0}(\Zbf)\,h_\ell(\Zbf)]-\E[\alpha_{\ell,0}(\Zbf)\,h_\ell(\Zbf)]=0$.

For the $m$-component of $\gamma$: $\partial_m\{X_k-m(\Zbf_k)\}[h_m]=-h_m(\Zbf_k)$, so by the same argument, $\partial_m\E[\gamma][h_m]=-\E[\alpha_{m,0}(\Zbf)\,h_m(\Zbf)]$ and the sum with~\eqref{eq:gateaux-m} vanishes.

\textit{Orthogonality with respect to $G$ and $S^T$.}
For admissible directions $H$, $\partial_{G}\E[\varphi^\star\given\Vbf][H]=0$ by Lemma~\ref{lem:G-cancellation}, since $Q^{I}$ satisfies~\eqref{eq:Q-cancellation}. At $G_0$, $S^T$ enters $\varphi^\star$ only through $Q^\star$ inside the martingale integral, whose conditional mean is zero for every small $t$ along $S^T_t=S^T_0+tH_S$, by predictability and the definition of these directions. Hence $\partial_{G}\E[\gamma][H]=\partial_{S^T}\E[\gamma][H_S]=0$. Joint perturbations of $G$ and $S^T$ are excluded from the admissible directions: their cross term, a product of the two errors as in~\eqref{eq:phi-double-robust}, is controlled by Assumption~\ref{ass:rates-L2}(iv).

\textit{Orthogonality with respect to $\alpha_\ell$ and $\alpha_m$.}
These enter $\gamma$ linearly. Since $\E[\varphi^\star_k\given\Zbf_k]=0$ at the truth and $\E[\nu_k\given\Zbf_k]=0$, we have $\partial_{\alpha_\bullet}\E[\gamma][h_{\alpha_\bullet}]=0$ for any square-integrable direction $h_{\alpha_\bullet}$, $\bullet\in\{\ell,m\}$.
\qed

\subsection{Proof of Lemma~\ref{lem:induced-smoothing-sharp}}\label{app:proof-induced-smoothing-sharp}

Write $\sigma_{ij}=\Gamma_n^{1/2}|X_{ij,0}|$ and $q_{ij}=q_{ij}(\cdot\given\Vbf_i,\Vbf_j)$. Conditionally on $(\Vbf_i,\Vbf_j)$, the substitution $t=\sigma_{ij}u$ gives
\[
\begin{aligned}
\E[\tilde g^*_{ij}-g^*_{ij}\given\Vbf_i,\Vbf_j]
&=X_{ij,0}\int\big\{\Phi(t/\sigma_{ij})-\I\{t\ge0\}\big\}\,q_{ij}(t)\,dt\\
&=X_{ij,0}\,\sigma_{ij}\int\big\{\Phi(u)-\I\{u\ge0\}\big\}\big\{q_{ij}(\sigma_{ij}u)-q_{ij}(0)\big\}\,du,
\end{aligned}
\]
where $q_{ij}(0)$ can be subtracted because $\Phi(u)-\I\{u\ge0\}$ is odd and integrable, and both sides vanish when $X_{ij,0}=0$. Hence
\[
\Gamma_n^{-1/2}\bigl|\E[\tilde g^*-g^*]\bigr|
\le\E\Big[X_{ij,0}^2\int\lvert\Phi(u)-\I\{u\ge0\}\rvert\,\lvert q_{ij}(\sigma_{ij}u)-q_{ij}(0)\rvert\,du\Big].
\]
As $\Gamma_n\to0$, the integrand tends to zero because $q_{ij}$ is continuous at $0$ by Assumption~\ref{ass:smoothing-rate}. It is bounded by $2L_0X_{ij,0}^2\lvert\Phi(u)-\I\{u\ge0\}\rvert$, which is integrable because $\E[X_{ij,0}^2]<\infty$ and $\int\lvert\Phi(u)-\I\{u\ge0\}\rvert\,du=2\phi(0)$. Dominated convergence on the product of the distribution of $(\Vbf_i,\Vbf_j)$ and Lebesgue measure in $u$ gives $\E[\tilde g^*-g^*]=o(\Gamma_n^{1/2})$.
\qed

\subsection{Proof of Lemma~\ref{lem:first-step}}\label{app:proof-first-step}

Fix $\beta_0$. Write $\omega_0=(\eta_0,\alpha_{\ell,0},\alpha_{m,0})$ and $\hat\omega^{(l)}=(\hat\eta^{(l)},\hat\alpha_\ell^{(l)},\hat\alpha_m^{(l)})$ for brevity. In the proofs below, $C$ denotes a constant that does not depend on $n$ and may change from line to line. The oracle U-statistic is
\[
U_n[\tilde\psi(\beta_0;\omega_0)]
=\frac{2}{n(n-1)}\sum_{1\le i<j\le n}\tilde\psi(\mathcal D_i,\mathcal D_j;\beta_0,\omega_0),
\]
and the cross-fitted version is
\[
U_n^{\rm cf}[\tilde\psi(\beta_0)]
=\frac{2}{n(n-1)}\sum_{l=1}^{L}\ \sum_{(i,j)\in\mathcal B_l}
\tilde\psi(\mathcal D_i,\mathcal D_j;\beta_0,\hat\omega^{(l)}).
\]
For each block $l$, let $\mathcal I_l=\bigcup_{(i,j)\in\mathcal B_l}\{i,j\}$,
$\mathcal I_{-l}=\{1,\dots,n\}\setminus \mathcal I_l$, and write
$\mathcal D_{-l}=\{\mathcal D_i:i\in\mathcal I_{-l}\}$.
By construction, $\hat\omega^{(l)}$ is measurable with respect to the $\sigma$-field generated by $\mathcal D_{-l}$ and auxiliary random variables independent of the data, while for each
$(i,j)\in\mathcal B_l$, $(\mathcal D_i,\mathcal D_j)$ is independent of this $\sigma$-field.

Define the block-wise kernel difference
\[
\Xi_{ij}^{(l)}
=\tilde\psi(\mathcal D_i,\mathcal D_j;\beta_0,\hat\omega^{(l)})
-\tilde\psi(\mathcal D_i,\mathcal D_j;\beta_0,\omega_0),
\qquad (i,j)\in\mathcal B_l,
\]
so that
$U_n^{\rm cf}[\tilde\psi(\beta_0)]-U_n[\tilde\psi(\beta_0;\omega_0)]
=\frac{2}{n(n-1)}\sum_{l=1}^{L}\sum_{(i,j)\in\mathcal B_l}\Xi_{ij}^{(l)}
= B_n + H_n$,
where
\begin{align*}
B_n
&=\frac{2}{n(n-1)}\sum_{l=1}^{L}\ \sum_{(i,j)\in\mathcal B_l}
\E\!\left[\Xi_{ij}^{(l)}\given \mathcal D_{-l}\right],\\
H_n &=\frac{2}{n(n-1)}\sum_{l=1}^{L}\ \sum_{(i,j)\in\mathcal B_l}
\Big[\Xi_{ij}^{(l)}-\E\{\Xi_{ij}^{(l)}\given \mathcal D_{-l}\}\Big].
\end{align*}

The following lemma bounds the second-order remainder of the base rank moment with a constant that does not depend on $\Gamma_n$.

\begin{lemma}\label{lem:g-remainder}
Let $\mathcal G(\ell,m)=\E[\tilde g(\mathcal D_i,\mathcal D_j;\beta_0,\ell,m)]$, $\delta_\ell=\ell-\ell_0$, $\delta_m=m-m_0$, and let $\dot{\mathcal G}[\delta_\ell,\delta_m]$ be the derivative of $r\mapsto\mathcal G(\ell_0+r\delta_\ell,m_0+r\delta_m)$ at $r=0$. Under Assumptions~\ref{ass:basic_id}(b), \ref{ass:rank-M2} and~\ref{ass:smoothing-rate}, if $\Gamma_n\le1$,
\[
\big|\mathcal G(\ell,m)-\mathcal G(\ell_0,m_0)-\dot{\mathcal G}[\delta_\ell,\delta_m]\big|
\le C\big(\|\delta_\ell\|_{2,w}^2+\|\delta_m\|_{2,w}^2\big),
\]
where $\|\cdot\|_{2,w}$ is the weighted norm in Assumption~\ref{ass:rates-L2} and $C$ depends only on $L_0$, $L_1$, $\beta_0$ and $\E\lvert\nu\rvert$.
\end{lemma}

\begin{proof}
By exchangeability, $\mathcal G(\ell,m)=\E[\tilde g^*_{ij}]$. Fix $i\neq j$ and condition on $(\Vbf_i,\Vbf_j)$. Write $s=\Gamma_n^{1/2}$, $x_0=X_{ij,0}$, $x=X_{ij}(m)$, $q=q_{ij}(\cdot\given\Vbf_i,\Vbf_j)$, $\bar Q(t)=\int_t^\infty q(u)\,du$, and $a=\delta_\ell(\Zbf_i)-\delta_\ell(\Zbf_j)-\beta_0\{\delta_m(\Zbf_i)-\delta_m(\Zbf_j)\}$, so that $e_j(\beta_0,\ell,m)-e_i(\beta_0,\ell,m)=D_{ij,0}+a$ and $x-x_0=-\{\delta_m(\Zbf_i)-\delta_m(\Zbf_j)\}$. Since $a$ and $x$ are functions of $(\Vbf_i,\Vbf_j)$, Assumption~\ref{ass:rank-M2} gives, for $x\neq0$,
\[
\E[\tilde g^*_{ij}\given\Vbf_i,\Vbf_j]
=x\,\E\Big[\Delta_i\Phi\Big(\frac{D_{ij,0}+a}{s|x|}\Big)\given\Vbf_i,\Vbf_j\Big]
=x\int\Phi\Big(\frac{t+a}{s|x|}\Big)q(t)\,dt=:F(a,x),
\]
and $F(a,0)=0$ by the convention after~\eqref{eqn:rank-smooth-moment}. Subscripts on $F$ denote partial derivatives. For $x\neq0$, write $\sigma=s|x|$ and $w=(t+a)/\sigma$. The derivatives $\partial_a\{x\Phi(w)\}=\mathrm{sgn}(x)\phi(w)/s$, $\partial_x\{x\Phi(w)\}=\Phi(w)-w\phi(w)$, $\partial_{ax}\{x\Phi(w)\}=w^2\phi(w)/\sigma$ and $\partial_{xx}\{x\Phi(w)\}=-w^3\phi(w)/x$ are bounded in $(t,a)$, locally uniformly in $x\neq0$, and $\int q\le1$, so dominated convergence justifies differentiation under the integral, and the substitution $t=\sigma u-a$ gives
\begin{align*}
F_a&=x\int\phi(u)\,q(\sigma u-a)\,du, & F_x&=\int\{\Phi(w)-w\phi(w)\}\,q(t)\,dt,\\
F_{ax}&=\int u^2\phi(u)\,q(\sigma u-a)\,du, & F_{xx}&=-s\,\mathrm{sgn}(x)\int u^3\phi(u)\,q(\sigma u-a)\,du .
\end{align*}
As $q\le L_0$ and $\int|u|^3\phi(u)\,du<2$, $|F_a|\le L_0|x|$, $|F_{ax}|\le L_0$ and $|F_{xx}|\le2L_0s$. As $\Phi(w)-w\phi(w)$ has derivative $w^2\phi(w)$, it increases from $0$ to $1$, so $0\le F_x\le1$. By Assumption~\ref{ass:smoothing-rate}, $F_a(\cdot,x)$ is Lipschitz with constant $L_1|x|$. As $(a',x)\to(a,0)$, $F_a(a',x)\to0$, and dominated convergence gives $F(a',x)/x\to\bar Q(-a)$ and $F_x(a',x)\to\bar Q(-a)$. Hence $F$ is continuously differentiable also across $x=0$, with $F_a(a,0)=0$ and $F_x(a,0)=\bar Q(-a)$, and $F_a(0,\cdot)$ and $F_x(0,\cdot)$ are Lipschitz with constants $L_0$ and $2L_0s$.

Along $(\ell_0+r\delta_\ell,m_0+r\delta_m)$, $(a,x)$ becomes $(ra,x_0+r(x-x_0))$ and the derivative of $F$ in $r$ is dominated by the integrable $L_0(|x_0|+|x-x_0|)|a|+|x-x_0|$, so $\dot{\mathcal G}[\delta_\ell,\delta_m]=\E\{F_a(0,x_0)a+F_x(0,x_0)(x-x_0)\}$ and the remainder in the lemma is the expectation of
\begin{multline*}
\{F(a,x)-F(0,x)-F_a(0,x)a\}+\{F_a(0,x)-F_a(0,x_0)\}a\\
+\{F(0,x)-F(0,x_0)-F_x(0,x_0)(x-x_0)\}.
\end{multline*}
The first term is at most $|x|\min(2L_0|a|,L_1a^2/2)\le L_1|x_0|a^2/2+2L_0|x-x_0||a|$, using $|x|\le|x_0|+|x-x_0|$. The second and third are at most $L_0|x-x_0||a|$ and $L_0s(x-x_0)^2\le L_0(x-x_0)^2$, as $s\le1$. Taking expectations, $\E[|x-x_0||a|]$ and $\E[(x-x_0)^2]$ are at most $C(\|\delta_\ell\|_2^2+\|\delta_m\|_2^2)$. Finally, since $|x_0|\le|\nu_i|+|\nu_j|$ with $\nu=X-m_0(\Zbf)$,
\[
\E\big[|X_{ij,0}|\{h(\Zbf_i)-h(\Zbf_j)\}^2\big]\le4\,\E\big[\{\E[|\nu|\given\Zbf]+\E|\nu|\}h(\Zbf)^2\big]\le C\|h\|_{2,w}^2,\qquad h\in\{\delta_\ell,\delta_m\},
\]
which bounds $\E[|x_0|a^2]$.
\end{proof}

\noindent
If $\E[|\nu|\given\Zbf]$ is bounded, $\|\cdot\|_{2,w}$ is equivalent to $\|\cdot\|_2$.

\medskip\noindent
\textbf{Step 1: the expectation-level term $B_n$.}
Let $\Psi(\omega)=\E[\tilde\psi(\mathcal D,\mathcal D';\beta_0,\omega)]$, so that $\E[\Xi_{ij}^{(l)}\given\mathcal D_{-l}]=\Psi(\hat\omega^{(l)})-\Psi(\omega_0)$ by conditional independence. Write $\hat\delta_\ell=\hat\ell^{(l)}-\ell_0$ and $\hat\delta_m=\hat m^{(l)}-m_0$, and take expectations over an independent draw given $\mathcal D_{-l}$. Since $\gamma$ is linear in $(\alpha_\ell,\alpha_m)$ and in $(\varphi^\star,X-m)$, $\E[\varphi_0\given\Zbf]=\E[\nu\given\Zbf]=0$, and $\hat\varphi^{(l)}-\varphi_0=\hat\rho^{(l)}-\rho_0-\hat\delta_\ell$,
\begin{multline*}
\E[\gamma(\mathcal D,\mathcal D';\hat\omega^{(l)})]
=\E[(\hat\alpha_\ell^{(l)}-\alpha_{\ell,0})(\hat\varphi^{(l)}-\varphi_0)]-\E[(\hat\alpha_m^{(l)}-\alpha_{m,0})\hat\delta_m]\\
+\E[\alpha_{\ell,0}(\hat\rho^{(l)}-\rho_0)]-\E[\alpha_{\ell,0}\hat\delta_\ell+\alpha_{m,0}\hat\delta_m],
\end{multline*}
while $\E[\gamma(\mathcal D,\mathcal D';\omega_0)]=0$. By Proposition~\ref{prop:rank-orthogonal}, the last term equals $-\dot{\mathcal G}[\hat\delta_\ell,\hat\delta_m]$. Hence Lemma~\ref{lem:g-remainder} and the Cauchy--Schwarz inequality give
\[
|\Psi(\hat\omega^{(l)})-\Psi(\omega_0)|
\le C\big\{\|\hat\delta_\ell\|_{2,w}^2+\|\hat\delta_m\|_{2,w}^2+r_n^{\alpha_\ell}r_n^\varphi+r_n^{\alpha_m}r_n^m\big\}+\bigl|\E[\alpha_{\ell,0}(\Zbf)\{\hat\rho^{(l)}-\rho_0\}]\bigr|,
\]
with $C$ not depending on $\Gamma_n$. Every term is $o_p(n^{-1/2})$ by Assumption~\ref{ass:rates-L2}, so $\sqrt n\,B_n=o_p(1)$.

\medskip\noindent
\textbf{Step 2: the centered empirical term $H_n$.}
Write $H_n=\sum_l\{2|\mathcal B_l|/n(n-1)\}H_{n,l}$, where $H_{n,l}$ is the average of $\Xi_{ij}^{(l)}-\E[\Xi_{ij}^{(l)}\given\mathcal D_{-l}]$ over $\mathcal B_l$. Conditionally on $\mathcal D_{-l}$, $H_{n,l}$ is a centered U-statistic of observations independent of $\mathcal D_{-l}$, with one sample for a triangular block and two samples for a rectangular one, and each index set has size proportional to $n$ because $K$ is fixed. Its conditional variance is therefore at most $C(n^{-1}v_{1,l}+n^{-2}v_{2,l})$, where $v_{1,l}$ is the variance of the first-order projection, summed over the two samples for a rectangular block, and $v_{2,l}$ is the variance of the kernel, both given $\mathcal D_{-l}$ and both at most a constant times $\E[(\Xi_{12}^{(l)})^2\given\mathcal D_{-l}]$, where $(1,2)$ denotes a generic pair in $\mathcal B_l$. By Chebyshev's inequality given $\mathcal D_{-l}$ and the fixed number of blocks, $\sqrt n\,H_n=o_p(1)$ follows once $\E[(\Xi_{12}^{(l)})^2\given\mathcal D_{-l}]=o_p(1)$ for each $l$. Since $\Xi_{12}^{(l)}$ is the sum of a base part and an adjustment part, it suffices that each has conditional second moment $o_p(1)$.

The adjustment part of $\Xi_{12}^{(l)}$ is $\{\hat\chi^{(l)}_1-\chi_{0,1}+\hat\chi^{(l)}_2-\chi_{0,2}\}/2$, with $\chi$ as in Assumption~\ref{ass:rates-L2}, so its conditional second moment is at most $(r_n^\chi)^2$. For the base part, write $X=X_{12}(\hat m^{(l)})$, $\sigma=\Gamma_n^{1/2}|X|$, $\sigma_0=\Gamma_n^{1/2}|X_{12,0}|$ and $a=\hat\delta_\ell(\Zbf_1)-\hat\delta_\ell(\Zbf_2)-\beta_0\{\hat\delta_m(\Zbf_1)-\hat\delta_m(\Zbf_2)\}$. Since $|\Phi(t/\sigma)-\I\{t>0\}|=\Phi(-|t|/\sigma)$ and the sub-densities of $D_{12,0}$ are bounded by $L_0$, the calculation in the proof of Lemma~\ref{lem:induced-smoothing} gives
\[
\E\Big[\frac{\Delta_1+\Delta_2}{2}\Big|\Phi\Big(\frac{D_{12,0}+a}{\sigma}\Big)-\Phi\Big(\frac{D_{12,0}}{\sigma_0}\Big)\Big|\given\Vbf_1,\Vbf_2\Big]\le L_0(|a|+\sigma+\sigma_0),
\]
and hence
\begin{multline}
\E\big[\{\tilde g(\mathcal D_1,\mathcal D_2;\beta_0,\hat\ell^{(l)},\hat m^{(l)})-\tilde g(\mathcal D_1,\mathcal D_2;\beta_0,\ell_0,m_0)\}^2\given\Vbf_1,\Vbf_2\big]\\
\le2(X-X_{12,0})^2+2X_{12,0}^2\min\{1,L_0(|a|+\sigma+\sigma_0)\}.
\label{eq:gtilde-L2}
\end{multline}
Given $\mathcal D_{-l}$, the first term has expectation at most $4\|\hat m^{(l)}-m_0\|_2^2$. In the second, $a$, $\sigma$ and $\sigma_0$ tend to zero in probability under the evaluation law when $\|\hat\ell^{(l)}-\ell_0\|_2+\|\hat m^{(l)}-m_0\|_2\to0$, since $\Gamma_n\to0$, and the term is bounded by the integrable $2X_{12,0}^2$. By Assumption~\ref{ass:rates-L2}, the subsequence characterization of convergence in probability and dominated convergence, both conditional expectations are $o_p(1)$.

Combining $\sqrt n\,B_n=o_p(1)$ and $\sqrt n\,H_n=o_p(1)$ proves the lemma.
\qed

\subsection{Proof of Proposition~\ref{prop:asymp-normal}}\label{app:proof-asymp-normal}

The kernel $\tilde\psi$ depends on $n$ through $\Gamma_n$, both in $\tilde g$ and in $(\alpha_{\ell,0},\alpha_{m,0})$, so the oracle U-statistic is a triangular array. Write $D_{ij}(\beta,\ell,m)=e_j(\beta,\ell,m)-e_i(\beta,\ell,m)$, $\delta_\ell=\ell-\ell_0$, and $\delta_m=m-m_0$. Then
\begin{equation}
D_{ij}(\beta,\ell,m)=D_{ij,0}+a_{ij},\qquad
a_{ij}=\delta_\ell(\Zbf_i)-\delta_\ell(\Zbf_j)-\beta\{\delta_m(\Zbf_i)-\delta_m(\Zbf_j)\}+(\beta-\beta_0)X_{ij,0},
\label{eq:D-shift}
\end{equation}
where $a_{ij}$ is a function of $(\Vbf_i,\Vbf_j)$, and $\Delta_i$ does not depend on $(\beta,\ell,m)$. Let $\bar q_{ij}(t)=\{q_{ij}(t\given\Vbf_i,\Vbf_j)+q_{ji}(-t\given\Vbf_j,\Vbf_i)\}/2$, so that $\E[(\Delta_i+\Delta_j)f(D_{ij,0})\given\Vbf_i,\Vbf_j]=2\int f(t)\,\bar q_{ij}(t)\,dt$ for bounded $f$. Kernels with $X_{ij}(m)=0$ follow the convention stated after~\eqref{eqn:rank-smooth-moment}. By Assumptions~\ref{ass:rank-M2} and~\ref{ass:smoothing-rate}, $\bar q_{ij}\le L_0$ and $|\bar q_{ij}(t)-\bar q_{ij}(0)|\le\min(2L_0,L_1|t|)$ almost surely. Since $\partial_\beta D_{ij}(\beta,\ell,m)=X_{ij}(m)$ and $\phi$ is even, differentiating~\eqref{eqn:rank-smooth-moment} and symmetrizing give, with $\sigma_{ij}=\Gamma_n^{1/2}|X_{ij}(m)|$ and $E_{ij}=D_{ij}(\beta,\ell,m)/\sigma_{ij}$,
\[
k_{ij}(\beta;\ell,m):=\partial_\beta\tilde g(\mathcal D_i,\mathcal D_j;\beta,\ell,m)
=\frac{\Delta_i+\Delta_j}{2}\,\phi(E_{ij})\frac{|X_{ij}(m)|}{\Gamma_n^{1/2}}\ \ge0 .
\]
By~\eqref{eq:D-shift} and the change of variables $u=(t+a_{ij})/\sigma_{ij}$,
\begin{equation}
\E[k_{ij}(\beta;\ell,m)\given\Vbf_i,\Vbf_j]=X_{ij}(m)^2\int\phi(u)\,\bar q_{ij}(\sigma_{ij}u-a_{ij})\,du .
\label{eq:jacobian-cond}
\end{equation}
For $m$ with $\|m-m_0\|_2<\infty$, $k_{ij}$ is bounded by the integrable $\phi(0)|X_{ij}(m)|/\Gamma_n^{1/2}$, so $\partial_\beta\E[\tilde g(\mathcal D_i,\mathcal D_j;\beta,\ell,m)]=\E[k_{ij}(\beta;\ell,m)]$.

\begin{lemma}\label{lem:jacobian}
Let $A=\E[X_{ij,0}^2\,q_{ij}(0\given\Vbf_i,\Vbf_j)]$ and $J_n(\beta;\ell,m)=\partial_\beta\E[\tilde g(\mathcal D_i,\mathcal D_j;\beta,\ell,m)]$. Under Assumptions~\ref{ass:basic_id}(b), \ref{ass:rank-M2} and~\ref{ass:smoothing-rate}, the following hold for every $M<\infty$.\\
(a) If $\|\ell_n-\ell_0\|_2+\|m_n-m_0\|_2\to0$, then $\sup_{|\beta-\beta_0|\le Mn^{-1/2}}|J_n(\beta;\ell_n,m_n)-A|\to0$.\\
(b) If, in addition, Assumption~\ref{ass:rates-L2} holds and $\liminf_n n\Gamma_n>0$,\hfil\break then $\sup_{|\beta-\beta_0|\le Mn^{-1/2}}\big|\partial_\beta U_n^{\rm cf}[\tilde\psi(\beta)]-A\big|\overset p\to0$.
\end{lemma}

\begin{proof}
\textit{(a)} By exchangeability, $A=\E[X_{ij,0}^2\,\bar q_{ij}(0)]$, and $J_n(\beta;\ell,m)$ is the expectation of~\eqref{eq:jacobian-cond}. With $(\delta_\ell,\delta_m)$ evaluated at $(\ell_n,m_n)$, $|a_{ij}|\le\bar a_{ij}:=|\delta_\ell(\Zbf_i)-\delta_\ell(\Zbf_j)|+(|\beta_0|+M)|\delta_m(\Zbf_i)-\delta_m(\Zbf_j)|+Mn^{-1/2}|X_{ij,0}|$ for $|\beta-\beta_0|\le Mn^{-1/2}$. Hence
\begin{multline*}
\sup_{|\beta-\beta_0|\le Mn^{-1/2}}|J_n(\beta;\ell_n,m_n)-A|
\le L_0\,\E|X_{ij}(m_n)^2-X_{ij,0}^2|\\
+\E\Big[X_{ij,0}^2\int\phi(u)\min\{2L_0,L_1(\sigma_{ij}|u|+\bar a_{ij})\}\,du\Big].
\end{multline*}
The first term vanishes because $\|m_n-m_0\|_2\to0$. In the second, $\sigma_{ij}\overset p\to0$ and $\bar a_{ij}\overset p\to0$, and the integrand is bounded by $2L_0X_{ij,0}^2\phi(u)$, so it vanishes by dominated convergence.

\textit{(b)} Since $\gamma$ does not depend on $\beta$, $\partial_\beta U_n^{\rm cf}[\tilde\psi(\beta)]$ is the average of $\bar k_l(\beta)=|\mathcal B_l|^{-1}\sum_{(i,j)\in\mathcal B_l}k_{ij}(\beta;\hat\ell^{(l)},\hat m^{(l)})$ over $l$ with weights $2|\mathcal B_l|/\{n(n-1)\}$, so it suffices to fix $l$. Condition on $\mathcal D_{-l}$, write $\beta_t=\beta_0+tn^{-1/2}$ and $X=X_{ij}(\hat m^{(l)})$, and let $c>0$ satisfy $n\Gamma_n\ge c$ for large $n$. Since $\E[\bar k_l(\beta_t)\given\mathcal D_{-l}]=J_n(\beta_t;\hat\ell^{(l)},\hat m^{(l)})$, (a) and the subsequence characterization of convergence in probability give $\sup_{|t|\le M}|\E[\bar k_l(\beta_t)\given\mathcal D_{-l}]-A|\overset p\to0$. For fixed $t$, $\bar k_l(\beta_t)$ is a one- or two-sample U-statistic of observations independent of $\mathcal D_{-l}$, indexed by sets of size proportional to $n$ because $K$ is fixed, so its conditional variance is at most a constant times $n^{-1}\E[k_{ij}(\beta_t)^2\given\mathcal D_{-l}]$. Since $\phi^2\le\phi(0)\phi$ and $\{(\Delta_i+\Delta_j)/2\}^2\le(\Delta_i+\Delta_j)/2$, the change of variables in~\eqref{eq:jacobian-cond} gives
\[
n^{-1}\E[k_{ij}(\beta_t)^2\given\Vbf_i,\Vbf_j]
\le\phi(0)X^2\min\Big\{\frac{\phi(0)}{n\Gamma_n},\frac{L_0|X|}{n\Gamma_n^{1/2}}\Big\}
\le\phi(0)X^2\min\Big\{\frac{\phi(0)}{c},\frac{L_0|X|}{(cn)^{1/2}}\Big\},
\]
whose expectation given $\mathcal D_{-l}$ vanishes in probability by dominated convergence and $\|\hat m^{(l)}-m_0\|_2\overset p\to0$. Hence $\bar k_l(\beta_t)-\E[\bar k_l(\beta_t)\given\mathcal D_{-l}]\overset p\to0$ for each $t$. For $|t|,|t'|\le M$, the arguments of $\phi$ in $k_{ij}(\beta_t)$ and $k_{ij}(\beta_{t'})$ differ by $|t-t'|(n\Gamma_n)^{-1/2}\le|t-t'|c^{-1/2}$, so
\[
|k_{ij}(\beta_t)-k_{ij}(\beta_{t'})|\le|t-t'|\,c^{-1/2}\,\frac{\Delta_i+\Delta_j}{2}\,\bar\phi(E_{ij})\frac{|X|}{\Gamma_n^{1/2}},
\]
where $E_{ij}$ is evaluated at $(\beta_0,\hat\ell^{(l)},\hat m^{(l)})$ and $\bar\phi(x)=\sup_{|y-x|\le Mc^{-1/2}}|\phi'(y)|$ is integrable. By~\eqref{eq:jacobian-cond} with $\phi$ replaced by $\bar\phi$,
\[
\E\Big[\sup_{|t|,|t'|\le M,\,|t-t'|\le\theta}|\bar k_l(\beta_t)-\bar k_l(\beta_{t'})|\given\mathcal D_{-l}\Big]\le\theta\,c^{-1/2}L_0\Big(\int\bar\phi(u)\,du\Big)\E[X^2\given\mathcal D_{-l}].
\]
For a grid of mesh $\theta$ on $[-M,M]$, $\sup_{|t|\le M}|\bar k_l(\beta_t)-A|$ is bounded by this oscillation, the maximum over the grid of $|\bar k_l(\beta_t)-\E[\bar k_l(\beta_t)\given\mathcal D_{-l}]|$, and $\sup_{|t|\le M}|\E[\bar k_l(\beta_t)\given\mathcal D_{-l}]-A|$. The last two vanish in probability, and the first is $O_p(\theta)$ with $\theta$ arbitrary.
\end{proof}

\medskip\noindent\textit{Proof of Proposition~\ref{prop:asymp-normal}.}
\textbf{Step 1: the oracle U-statistic.} Write $\tilde\psi_n=\tilde\psi(\cdot,\cdot;\beta_0,\omega_0)$, $\Psi_n=\E[\tilde\psi_n(\mathcal D_i,\mathcal D_j)]$, and $\zeta_n(\mathcal D_i)=\E[\tilde\psi_n(\mathcal D_i,\mathcal D_j)\given\mathcal D_i]-\Psi_n$. The Hoeffding decomposition \citep{hoeffding_class_1948} gives
\[
U_n[\tilde\psi(\beta_0;\omega_0)]-\Psi_n=\frac2n\sum_{i=1}^n\zeta_n(\mathcal D_i)+R_n,\qquad
\E[R_n^2]\le\frac{2\,\E[\tilde\psi_n^2]}{n(n-1)}\le\frac{2\,\E[b^2]}{n(n-1)},
\]
because $R_n$ is a degenerate U-statistic whose kernel has variance at most $\E[\tilde\psi_n^2]$, and $b$ does not depend on $n$ (Assumption~\ref{ass:score-regularity}(a)). Hence $\sqrt n\,R_n=O_p(n^{-1/2})$. Since $\E[\gamma(\mathcal D_i,\mathcal D_j;\eta_0,\alpha_{\ell,0},\alpha_{m,0})]=0$ and $\E[g^*]=0$ (Lemma~\ref{lem:gehan-unbiased}), $\sqrt n\,\Psi_n=\sqrt n\,\E[\tilde g^*-g^*]=o(1)$ by Lemma~\ref{lem:induced-smoothing-sharp} and $\Gamma_n=O(n^{-1})$.

Next, $\tilde\psi_n$ has an almost sure limit. Since $\prob(\Delta_i=1,D_{ij,0}=0)=0$ by Assumption~\ref{ass:rank-M2}, $\tilde g(\cdot,\cdot;\beta_0,\ell_0,m_0)\to g=\tfrac12(g^*_{ij}+g^*_{ji})$ almost surely. For the weights~\eqref{eqn:pairwise-weights}, $W_{ij,\ell}\to0$ pointwise, but integrating $D_{ij,0}$ out first as in~\eqref{eq:jacobian-cond} gives $\E[W_{ij,\ell}\given\Vbf_i,\Vbf_j]=X_{ij,0}\int\phi(u)\,\bar q_{ij}(\sigma_{ij}u)\,du\to X_{ij,0}\,\bar q_{ij}(0)$ almost surely, by the continuity of $\bar q_{ij}$ at $0$. In $\E[W_{ij,m}\given\Vbf_i,\Vbf_j]$, the term with $\beta_0\,\mathrm{sgn}(X_{ij,0})/\Gamma_n^{1/2}$ converges in the same way, the term with $\phi(E_{ij,0})E_{ij,0}$ is at most $L_0\sigma_{ij}$ in absolute value, and the term with $\Phi(E_{ij,0})$ converges because $\Phi(E_{ij,0})\to\I\{D_{ij,0}>0\}$ almost surely. These conditional means are bounded by $C(1+|X_{ij,0}|)$, so dominated convergence given $\Zbf_i$ shows that $\alpha_{\bullet,0}(\zbf)=2\E[\E\{W_{ij,\bullet}\given\Vbf_i,\Vbf_j\}\given\Zbf_i=\zbf]$ converges for almost every $\zbf$ to a limit $\alpha_{\bullet,\infty}(\zbf)$, $\bullet\in\{\ell,m\}$. Hence $\tilde\psi_n\to\psi_\infty=g+\gamma(\cdot,\cdot;\eta_0,\alpha_{\ell,\infty},\alpha_{m,\infty})$ almost surely. As $|\tilde\psi_n|\le b$ for all $n$, $\tilde\psi_n\to\psi_\infty$ in $L_2$, so $\E[\psi_\infty]=\lim_n\Psi_n=0$ and, by Jensen's inequality, $\zeta_n\to\zeta$ in $L_2$ with $\zeta(\mathcal D_i)=\E[\psi_\infty(\mathcal D_i,\mathcal D_j)\given\mathcal D_i]$. In particular, the limit $V=\lim_n4\V\{\zeta_n(\mathcal D)\}=4\V\{\zeta(\mathcal D)\}$ exists. The summands $\zeta_n(\mathcal D_i)-\zeta(\mathcal D_i)$ are i.i.d.\ with mean zero, so
\[
\E\Big[\Big\{\frac{2}{\sqrt n}\sum_{i=1}^n\{\zeta_n(\mathcal D_i)-\zeta(\mathcal D_i)\}\Big\}^2\Big]=4\,\E[\{\zeta_n(\mathcal D)-\zeta(\mathcal D)\}^2]\to0,
\]
and the central limit theorem for the fixed function $\zeta$ gives $\sqrt n\,U_n[\tilde\psi(\beta_0;\omega_0)]\overset d\to N(0,V)$. By Lemma~\ref{lem:first-step}, $Z_n:=\sqrt n\,U_n^{\rm cf}[\tilde\psi(\beta_0)]\overset d\to N(0,V)$.

\medskip\noindent
\textbf{Step 2: linearization.} Since $\gamma$ does not depend on $\beta$ and $k_{ij}\ge0$, $\beta\mapsto U_n^{\rm cf}[\tilde\psi(\beta)]$ is continuously differentiable and nondecreasing. It is strictly increasing unless $(\Delta_i+\Delta_j)X_{ij}(\hat m^{(l)})=0$ for all pairs, so its root is unique when it exists. For $t\in\mathbb R$,
\[
\sqrt n\,U_n^{\rm cf}[\tilde\psi(\beta_0+tn^{-1/2})]=Z_n+\int_0^t\partial_\beta U_n^{\rm cf}[\tilde\psi(\beta_0+sn^{-1/2})]\,ds,
\]
so Lemma~\ref{lem:jacobian}(b) gives $\sup_{|t|\le M}|\sqrt n\,U_n^{\rm cf}[\tilde\psi(\beta_0+tn^{-1/2})]-Z_n-At|\overset p\to0$ for every $M$. Since $Z_n$ is tight, for every $\kappa>0$ there is $M$ such that, for large $n$, $|Z_n|\le AM/2$ and this supremum is below $AM/4$ with probability at least $1-\kappa$. On this event, the nondecreasing map $t\mapsto\sqrt n\,U_n^{\rm cf}[\tilde\psi(\beta_0+tn^{-1/2})]$ is negative at $t=-M$ and positive at $t=M$, so the root $\hat\beta$ exists, $\sqrt n|\hat\beta-\beta_0|<M$, and $|Z_n+A\sqrt n(\hat\beta-\beta_0)|$ is bounded by the supremum. Since $\kappa$ is arbitrary and $A>0$ by Assumption~\ref{ass:score-regularity}(b),
\[
\sqrt n(\hat\beta-\beta_0)=-A^{-1}Z_n+o_p(1)\ \overset d\to\ N\!\left(0,\frac{V}{A^2}\right).
\]
\qed

\subsection{Consistency of the variance estimators}\label{app:variance}

Let $\chi_k(\omega)=\alpha_\ell(\Zbf_k)\varphi^\star_k(\ell,G,S^T)+\alpha_m(\Zbf_k)\{X_k-m(\Zbf_k)\}$ as in Assumption~\ref{ass:rates-L2}, so that $\gamma(\mathcal D_i,\mathcal D_j;\omega)=\{\chi_i(\omega)+\chi_j(\omega)\}/2$, and write $\chi_{0,k}=\chi_k(\omega_0)$. Since $\chi$ contains the product $\alpha_\ell\varphi^\star$, the rate $r_n^\chi$ in Assumption~\ref{ass:rates-L2} is not controlled by $r_n^{\alpha_\ell}$ and $r_n^\varphi$ alone.

\begin{lemma}\label{lem:variance}
Under the conditions of Proposition~\ref{prop:asymp-normal}, $\hat A\overset p\to A$. Let $\hat\chi^{(l)}_k$ be $\chi_k$ at the block-$l$ nuisance estimates used in $\hat V$. If
\begin{equation}
\max_l\,|\mathcal I_l|^{-1}\sum_{k\in\mathcal I_l}\big(\hat\chi^{(l)}_k-\chi_{0,k}\big)^2=o_p(1),
\label{eq:chi-insample}
\end{equation}
then $\hat V\overset p\to V$. If these estimates depend on the data only through $\mathcal D_{-l}$, as $\hat\omega^{(l)}$ does, \eqref{eq:chi-insample} follows from $r_n^\chi=o_p(1)$ and Markov's inequality. The refit of $(\hat\alpha_\ell,\hat\alpha_m)$ at $\hat\beta$ depends on $\mathcal I_l$ only through $\hat\beta$. Let $\hat\chi^{(l)}_k(\beta)$ be $\hat\chi^{(l)}_k$ with the refit at a fixed $\beta$, which depends on the data only through $\mathcal D_{-l}$. Then \eqref{eq:chi-insample} holds for the refit at $\hat\beta$ if, for every $M<\infty$, $\max_l\E[\sup_{|\beta-\beta_0|\le Mn^{-1/2}}\{\hat\chi^{(l)}(\beta)-\chi_0\}^2]=o_p(1)$, with the expectation taken as in Assumption~\ref{ass:rates-L2}. This is a locally uniform version of $r_n^\chi=o_p(1)$ for the refit.
\end{lemma}

\begin{proof}
Since $\gamma$ does not depend on $\beta$, $\hat A=\partial_\beta U_n^{\rm cf}[\tilde\psi(\beta)]$ at $\beta=\hat\beta$, so $\hat A\overset p\to A$ by Lemma~\ref{lem:jacobian}(b) and $\sqrt n(\hat\beta-\beta_0)=O_p(1)$.

For $\hat V$, let $\hat\psi_{ij}$ be its summand, $\psi_{ij}=\tilde\psi_n(\mathcal D_i,\mathcal D_j)$, $\hat h_i=(n-1)^{-1}\sum_{j\neq i}\hat\psi_{ij}$, $h_i=(n-1)^{-1}\sum_{j\neq i}\psi_{ij}$, and $\zeta_i=\zeta(\mathcal D_i)$ with $\zeta$ from Step~1 of the proof of Proposition~\ref{prop:asymp-normal}. Since $n^{-1}\sum_i\zeta_i^2\overset p\to V/4$ and $|\hat V^{1/2}-(4n^{-1}\sum_i\zeta_i^2)^{1/2}|\le2\{n^{-1}\sum_i(\hat h_i-\zeta_i)^2\}^{1/2}$, it suffices that $n^{-1}\sum_i(\hat h_i-\zeta_i)^2\overset p\to0$. Given $\mathcal D_i$, $h_i$ averages $n-1$ i.i.d.\ terms with mean $\zeta_n(\mathcal D_i)+\Psi_n$, so $\E[(h_i-\zeta_i)^2]\le3\E[b^2]/(n-1)+3\E[\{\zeta_n(\mathcal D)-\zeta(\mathcal D)\}^2]+3\Psi_n^2\to0$ by Step~1. By the Cauchy--Schwarz inequality, $n^{-1}\sum_i(\hat h_i-h_i)^2$ is at most the average of $(\hat\psi_{ij}-\psi_{ij})^2$ over all pairs, so it suffices that its average over each block $\mathcal B_l$ vanishes in probability. For $(i,j)\in\mathcal B_l$, write $X=X_{ij}(\hat m^{(l)})$ and split $\hat\psi_{ij}-\psi_{ij}$ into three parts.

\textit{The change in $\beta$.} On the event $\sqrt n|\hat\beta-\beta_0|\le M$, whose probability is arbitrarily close to one for large $M$, the arguments of $\Phi$ in $\tilde g$ at $\hat\beta$ and at $\beta_0$ differ by at most $M(n\Gamma_n)^{-1/2}$. As $\Phi$ is monotone and $\Phi(-x)=1-\Phi(x)$, the difference of $\tilde g$ is at most
\[
S_{ij}=|X|\,\frac{\Delta_i+\Delta_j}{2}\,\big\{\Phi\big(E_{ij}+M(n\Gamma_n)^{-1/2}\big)-\Phi\big(E_{ij}-M(n\Gamma_n)^{-1/2}\big)\big\},
\]
with $E_{ij}$ evaluated at $(\beta_0,\hat\ell^{(l)},\hat m^{(l)})$. Since $\int\{\Phi(u+c)-\Phi(u-c)\}du=2c$, the change of variables in~\eqref{eq:jacobian-cond} gives $\E[S_{ij}^2\given\Vbf_i,\Vbf_j]\le X^2\min\{1,2ML_0|X|n^{-1/2}\}$, whose expectation given $\mathcal D_{-l}$ vanishes in probability by dominated convergence and $\|\hat m^{(l)}-m_0\|_2\overset p\to0$.

\textit{The change in $(\ell,m)$ at $\beta_0$.} By~\eqref{eq:gtilde-L2} and the argument following it in the proof of Lemma~\ref{lem:first-step}, the conditional mean of its square given $\mathcal D_{-l}$ is $o_p(1)$.

\textit{The adjustment.} Since $\gamma$ averages $\chi_i$ and $\chi_j$, the adjustment parts $\hat\gamma_{ij}$ and $\gamma_{ij}$ of $\hat\psi_{ij}$ and $\psi_{ij}$ satisfy $(\hat\gamma_{ij}-\gamma_{ij})^2\le\{(\hat\chi^{(l)}_i-\chi_{0,i})^2+(\hat\chi^{(l)}_j-\chi_{0,j})^2\}/2$. Each $k\in\mathcal I_l$ appears in at most a fraction $C/|\mathcal I_l|$ of the pairs in $\mathcal B_l$, so the block average of this bound is at most $C$ times the left-hand side of~\eqref{eq:chi-insample}. For the refit at $\hat\beta$ under the locally uniform condition, on the event $\sqrt n|\hat\beta-\beta_0|\le M$ this left-hand side is at most $\max_l|\mathcal I_l|^{-1}\sum_{k\in\mathcal I_l}\sup_{|\beta-\beta_0|\le Mn^{-1/2}}\{\hat\chi^{(l)}_k(\beta)-\chi_{0,k}\}^2$, which is $o_p(1)$ by Markov's inequality given $\mathcal D_{-l}$. The event has probability arbitrarily close to one for large $M$ by Proposition~\ref{prop:asymp-normal}.
\end{proof}

\end{document}